\documentstyle[11pt,aaspp4,flushrt]{article} 
\tighten 
\begin{document}
\newcommand{\cmtwo}{cm$^{-2}$}
\newcommand{\degpoint}{\mbox{$^\circ\mskip-7.0mu.\,$}}
\newcommand{\kms}{\,km~s$^{-1}$}      
\newcommand{\minpoint}{\mbox{$'\mskip-4.7mu.\mskip0.8mu$}}
\newcommand{\mv}{\mbox{$m_{_V}$}}
\newcommand{\Mv}{\mbox{$M_{_V}$}}
\newcommand{\peryr}{\mbox{$\>\rm yr^{-1}$}}
\newcommand{\secpoint}{\mbox{$''\mskip-7.6mu.\,$}}
\newcommand{\sqdeg}{\mbox{${\rm deg}^2$}}
\newcommand{\squig}{\sim\!\!}
\newcommand{\subsun}{\mbox{$_{\odot}$}}
\newcommand{\et}{et al.~}
\newcommand{\cf}{c.f.~}
\newcommand{\eg}{e.g.~}

\def\ltsima{$\; \buildrel < \over \sim \;$}
\def\simlt{\lower.5ex\hbox{\ltsima}}
\def\gtsima{$\; \buildrel > \over \sim \;$}
\def\simgt{\lower.5ex\hbox{\gtsima}}
\def\arcs{$''~$}
\def\arcm{$'~$}
\def\kmm{{\sc kmm}}

\title {THE AGES AND ABUNDANCES OF A LARGE SAMPLE OF M87 GLOBULAR CLUSTERS\altaffilmark{1}}
\author{\sc Judith G. Cohen\altaffilmark{2}, 
John P. Blakeslee\altaffilmark{2}, and 
Anton Ryzhov\altaffilmark{2,}\altaffilmark{3}}

\altaffiltext{1}{Based in large part on observations obtained at the
W.M. Keck Observatory, which is operated jointly by the California 
Institute of Technology and the University of California}
\altaffiltext{2}{Palomar Observatory, Mail Stop 105-24,
California Institute of Technology, Pasadena, California 91125}
\altaffiltext{3}{Current address: Department of Physics, Box 351560,
University of Washington, Seattle, Washington 98195-1560}

\begin{abstract}

A subset of
150 globular clusters (GCs) in M87 has been selected for abundance and 
age determinations from the sample of Paper I (Cohen \& Ryzhov 1997).  
This has been
done solely on the basis of the signal-to-noise ratios of the spectra.
Indices that measure the strength of the
strongest spectral features were determined for the M87 GCs and
from new data for twelve galactic GCs.  Combining the new and existing data
for the galactic GCs and comparing the $(U-R)$ colors and the
line indices gave qualitative indications for the ages and
abundances of the M87 GC system.

Quantitative results, which confirm and extend the qualitative
ones, were obtained by applying
the Worthey (1994) models for the integrated light of stellar
systems of a single age, calibrated by 
observations of galactic
globular clusters, to deduce abundances and ages for the objects
in our sample.

We find that the M87 GCs span a wide range in metallicity, from
very metal poor to somewhat above solar metallicity.  The mean
[Fe/H] of $-0.95$ dex
is higher than that of the galactic GC system, and there is a metal
rich tail that reaches to higher [Fe/H] than one finds among the
galactic GCs.   Excluding the very metal rich tail, there 
is marginal evidence for a bimodal distribution
over the single one at the 89\% significance level.
The two ``subpopulations'' in this model are
located at $-$1.3 and $-$0.7 dex and contain 40\% and 60\% of the total,
respectively.  The dispersion in [Fe/H] for each of the model subpopulations
is $\sigma{\,=\,}$0.3~dex.

The mean metallicity of the M87 GC system is about a factor of four 
lower than that of the M87 stellar halo at a fixed projected radius $R$.  
The metallicity inferred from the X-ray studies is similar to that of the
M87 stellar halo, not to that of the M87 GC system.

We infer the relative abundances of Na, Mg, and Fe in the M87 GCs
from the strength of their spectral features.  The behavior
of these elements 
between the metal rich and metal poor M87 GCs is similar to that shown
by the galactic GCs and by halo stars in the Galaxy.  The 
pattern of chemical
evolution in these disparate old stellar systems is, as far as we can tell,
identical.

Superposed on a very large dispersion in abundance at all $R$,
there is a small but real radial gradient in the mean abundance
of the M87 GCs with $R$, but no detectable change
in the $H_{\beta}$ index with $R$.

We obtain a median age for the M87 GC system of 13 Gyr, similar to that
found for the galactic GCs.  The dispersion about that value
($\sigma$ = 2 Gyr) is small.

\end{abstract}


\section{INTRODUCTION}

This is the second in a series of papers on the M87 globular cluster 
system.  In Paper I (Cohen \& Ryzhov 1997) 
a sample of M87 GCs was isolated from 
the Strom et al (1981) photographic survey for globular cluster 
candidates in M87, supplemented by additional candidates in the 
core of M87 identified in Paper~I, and  
radial velocities were presented for about 300 
of these candidates from LRIS/Keck spectra in the slitmask mode.
Based on their radial velocities, 205 objects from this sample were 
selected that are believed to be bona fide M87 GCs.  In the first paper we 
explored the dynamics of this rich GC system, focusing on such issues 
as the spatial distribution of the M87 GCs and of the M87 halo light, 
on the velocity dispersion as a function of projected radius, on the deprojections needed to derive the enclosed $M(r)$, and on the 
$M/L$ ratio 
as a function of $r$. 

In this paper we turn to the inference of abundance 
and age for the M87 GCs.  
There have been many efforts along these lines in the past, but
because these objects are so faint, most such studies have involved 
broad or narrow band photometry of various kinds (Strom et al 1981, 
Cohen 1988, Couture et al 1990, Lee \& Geisler 1993). 
This is a tradeoff of desired sample size with desired accuracy of 
and detail in the results and also ease of calibration of the 
results.  Until the advent of very large ground based telescopes, 
the photometric surveys were clearly at an advantage.  

Imaging with HST has offered unprecedented photometric accuracy for 
faint objects such as the M87 GCs, and although the field of view is very 
small (meaning that spatial gradients within the halo of M87 cannot 
be followed), many interesting issues have been pursued (Whitmore et al 1995,
Elson and Santiago 1996a,b).  The major new result 
that has emerged from this work is a strong demonstration
of a bimodal color distribution, first suggested in the
ground based work of Lee \& Geisler (1993).  
We will attempt to verify that here and to discern its origin.

Spectroscopic efforts began with Racine et al (1978) and Hanes \& 
Brodie (1986) who succeeded in reaching only 6 of the brightest 
M87 GCs at very low resolution.  As instrumentation improved, 
efforts continued at Palomar and at the MMT (Mould et al 1987, 
Mould et al 1990, Brodie \& Huchra 1991) at moderate spectral 
resolution to produce a total sample of about 45 observed
objects believed to be M87 GCs, but  
many of their spectra are of quite low S/N and suitable only for 
radial velocity determinations.

Our work takes advantage of the large collecting area of the
the Keck I 10-m Telescope and of an efficient multi-object 
spectrograph, LRIS (Oke et al 1995).  We have assembled a 
statistically significant sample of object with 
high precision spectra, and thus will be able to deduce relatively 
precise abundances and ages for the M87 GCs.

\section{THE SAMPLE OF M87 GLOBULAR CLUSTERS}

The sample of objects under consideration here is the sample of
205 objects believed to be M87
GCs isolated in Paper~I.  We use only the spectra taken in 
the 5200\AA~region; we do not use the spectra taken in
the 8500\AA~region due to the excessive detector noise present in the
LRIS at the time (winter 1994) that the redder spectra were taken.

An accurate $v_r$ can be obtained from noisier spectra than are 
required for accurate abundance analyses because of the multiplexing 
of lines in cross correlation  $v_r$ measurements and the 
characterization of the entire spectrum by only a single parameter.
To avoid introducing excessive noise into the measurements, correlations,
and analysis presented below, 
we restrict the sample based on the brightness of an object as
viewed through the LRIS slit in these LRIS multi-slit spectra. 
The final sample consists of those objects whose continuum
level is such (with respect to the background 
from the night sky and from the halo of M87 itself seen in
the particular slitlet of a multi-slit mask corresponding to
the object) that a S/N ratio
of 22 or greater per pixel (1.24 \AA/pixel) is achieved in the continuum.
This is calculated over the region from 5450 to 5500\AA~assuming
that the spectrum is summed over 8 pixels along the slit, where
the scale is 4.7 pixels/arc-sec.  (The actual height along the slit 
extracted during the reduction of each spectrum
varied due to variations in the seeing.)
Since the readout noise is small, we evaluate this for the sum of 
all spectra of a given object taken with a given slitmask here, 
normally two or three 3000 sec exposures.  
We ignore the effects of cosmic rays in this calculation. 
For future reference this value is denoted as ``$QSNR$''. 

An object may get into the sample on the basis of a single 
spectrum with $QSNR \ge 22$.  If there are two such spectra
for an object, they provide
independent measurements.  In two cases, sums of two spectra 
of a particular object taken with different slitmasks
met the requirement for $QSNR$ even though  
individually each is too noisy.

This test is still not sufficient to
produce a sample whose minimum S/N is uniform,
due to the variation in the placement of objects within
the individual slits.
If an object is very close to the edge of a slitlet in the 
slitmask, sky subtraction is much harder, and the S/N ratio, 
particularly in areas with strong night sky emission, 
may be adversely affected. 
However, as long as our selection criteria are independent 
of the strength of the absorption features in the spectra, they 
should suffice for present purposes. 

Two objects (Strom 714 and Strom 1067) were rejected because the
appearance of their spectra (not in terms of the spectral 
features, but rather the 1d spatial profile along the slit)
indicated that the objects were not single point sources, but
rather slightly extended, or a close pair of point sources.

We end up with a sample for the abundance analysis containing 
162 spectra of 150 M87 GCs.  The fraction of
objects rejected through failure to meet the minimum
S/N ratio criterion is small, ${\approx}6\%$.
%
%
%

\section{DATA ANALYSIS} 

\subsection {Adopted Indices}

There are many systems of indices designed to characterize the
strength of the most prominent absorption features in stellar spectra.
We have adopted the Lick indices as defined by Burstein et al (1984),
Faber et al (1985), and Gorgas et al (1993)
for our work on the M87 GCs 
because of the extensive database in the literature in this system.  
Measured indices for large numbers of galactic stars are described in
Faber et al (1985) and Gorgas et al (1993).  Indices in this system are 
given for 17 galactic GCs and 19 GCs in M31 by Burstein et al (1984).  
Brodie \& 
Huchra (1990) present indices in the Lick system for 
additional M31 and galactic globular clusters,
while Burstein et al (1984) present indices for a sample of nearby galaxies.  Worthey (1994) has carried out extensive modeling combining the 
fluxes predicted by grids of theoretical stellar atmospheres 
(i.e. Kurucz 1993) with a set of evolutionary tracks to predict 
semi-empirical indices for the integrated light of single age burst 
models of composite stellar systems of varying metallicity.  
Chavez et al (1995) also attempted to calibrate the Lick indices for 
the Mg region.

Lick indices are not defined using fluxed spectra.
The original spectra were all taken with the IDS (Robinson \& Wampler 1972), 
a detector in use at Lick for many years, and were 
normalized to observations of a quartz lamp.  It is thus 
difficult to reproduce at present the continuum slope
of this data without additional information.  

Our M87 GC spectra are given as a function of DN (2 $e^{-1}$/DN) versus
$\lambda$ rather than flux as a function of $\lambda$.  
They can not easily nor accurately be converted into flux 
because of variable light loss in the LRIS slits due to 
alignment errors, and because some of the nights during which
spectra were obtained were not photometric.  In addition,
as described in Paper~I,
two gratings with the same dispersion but different blaze angles
were used during the 15 month period over which these observations
were made.

One therefore has to be careful of the effect of global continuum shape
on the measured line indices.  Under such circumstances
narrow indices should be more reproducible, and we
have therefore favored the narrower indices whenever possible.  
Thus of the three Mg indices defined in the Lick system, 
we use Mg $b$.  It measures only the
Mg triplet, while Mg$_1$ and Mg$_2$ also include the MgH band.
The M87 GCs are not of such a late type that the MgH band is
very strong, so use of the Mg $b$ index should suffice.
Measurements of the Mg$_1$ and Mg$_2$ indices show a systematic offset with
respect to the Lick measurements, presumably due to differences
in continuum slope.
All trends described here with respect to Mg $b$ are also obeyed
by the Mg$_1$ and Mg$_2$ indices. 

Worthey \& Ottaviani (1997) discuss transformation of measured indices
to the Lick system, emphasizing spectral resolution and velocity
dispersion effects.  However, since our spectral resolution is 8\AA,
while that of the Lick IDS was 8 -- 10 \AA~depending somewhat on
wavelength, we do not need to smooth our spectra to match.
No correction for internal velocity dispersion was made either.  Unlike the
case of galaxies, the velocity dispersion inside GCs is small enough
that this is not necessary.

Since there was no Lick index for $H_{\alpha}$, we
defined one.  The bandpasses for the adopted indices are listed in Table 1.
%
%

\subsection {Data Analysis}

As described in Paper~I, all the spectra were reduced using Figaro 
scripts written by A.Ryzhov.  It became clear that some improvements 
could be made particularly in the area of cosmic ray removal and 
extraction of the final spectrum. About 80 spectra were re-reduced 
using improved scripts and 
a procedure which required more manual intervention but produced 
spectra with significantly reduced  noise.  The spectra selected
for re-reduction were those at the faint end of the sample and those
just below the cutoff in the hopes that a re-reduction would get
them into the sample.

The $v_r$ adopted (all the bandpasses used to compute the
indices are of course shifted by the
appropriate factor for the $v_r$ of each object) are those of
Paper~I.  In the course of the re-reduction of the spectra, 
we noticed 5 objects whose measured
$v_r$ appeared to be significantly ($\ge 100$ \kms) different from
the value published in Paper~I.  Also the spectrum of one object 
was not correctly reduced initially.  It was not included in Paper~I,
but now has a definite $v_r$ indicating membership in the M87 GC system
and a S/N ratio such that it is included in the abundance sample.
The corrected $v_r$ for these five objects and the new $v_r$ for 
Strom 682 are given in Appendix A.
%
%

Two of the M87 GCs (Strom 519 and Strom 750)
showed emission at the [OIII] doublet at 4959 and 5007\AA.
The emission was clearly extended several arc-sec beyond 
the cluster in the case of Strom 750, and emission
at $H_{\alpha}$ was also seen.  This emission is assumed to 
originate in the filaments near the nucleus of M87 and is
subsequently ignored.

These are faint objects, and even with the best possible algorithm
for the removal
of the night sky features, there are often strong residuals of
the NaD lines in the terrestrial atmosphere.  Since the radial
velocity of an object in M87 is by definition at least 250 \kms,
and $v_r$ for M87 itself is 1277 \kms, each of the spectra has
been inspected by eye, and when necessary, an interpolation of
the continuum has been made to remove the residual night sky emission.
Such a procedure was necessary and has been done for almost
all of these spectra.

Figure 1 shows a comparison of the spectrum of a relatively high
(Strom 1010, [Fe/H]$_Z = -0.2$ dex)
and relatively low metallicity GC (Strom 910, [Fe/H]$_Z = -2.2$ dex ) in M87
over the regime from 4600 to 6650\AA. 
Both of these objects
have $QSNR \approx 40$.
The residuals of the three strongest
night sky features in this spectral region (at 5577, 5891 and 6300 \AA)
have been removed via interpolation.
The strongest features are identified.
Note the relative prominence of the Balmer lines in the low metallicity
object and the much stronger metallic absorption features in the
high metallicity M87 GC.  


Because these are spectra taken with a multi-slit instrument,
the wavelength range of the spectrum of a particular object
depends on the position of the object in the telescope focal plane.  This means
that the reddest and bluest features analyzed ($H_{\beta}$ and $H_{\alpha}$)
may not be present on all the spectra. 
Each spectrum was checked to make sure there would be no problems
with either of the Balmer line indices due to a bandpass slipping beyond
the range of the valid data.

\subsection {Measured Indices}

Table 2 contains the final list of measured indices for the
M87 GCs together with their $QSNR$ values.
%
%

The errors for the indices have been calculated assuming Poisson statistics,
taking into account the number of pixels averaged to form the
numerator and denominator of the index, and also the change in mean signal
level (in DN) of the spectra as a function of wavelength.  
This last is a factor of 2.5 from
the lowest mean signal level in DN at the blue end of the $H_{\beta}$ index
to the higher signal levels prevailing at the redder indices.  
Assuming $QSNR = 22$
predicts errors of 0.5\AA~for the $H_{\beta}$ index, the same 
for the Mg $b$ index, and 0.3\AA~for the $H_{\alpha}$ index. 
The predicted error for the TiO$_1$ index is smaller, as many more pixels
are averaged in each of its bandpasses.
These calculated values do not take into account
cosmic rays or the systematic problems of
matching the global continuum slope discussed above.   Also the
above were predicated on the smooth base level of the night sky and
do not include the individual strong resolved features such as the
NaD lines that fall within the redder indices.

The mean of the absolute value of the
differences in the measured indices for the 12 M87 GCs with
multiple spectra is 0.5\AA~ or less for the five indices we use.
This ignores three outliers among the 60 cases.

A foreground extinction of $E(B{-}V) = 0.022$ mag to M87 (Burstein \&
Heiles 1984) and the reddening law of Reed et al (1988)
are adopted for the M87 GCs.

\section{CALIBRATION VIA THE GALACTIC GLOBULAR CLUSTERS}

As a calibration for the M87 globular clusters, we rely on
observations of galactic globular clusters.  We have scanned
the cores of 
12 galactic globular clusters covering a wide range in metallicity
using the same instrumental configuration as was used for 
our M87 GC observations.  The scans are typically 90 arc-sec
in length and we use a 3 arc-min long slit, so the stellar population
within a cluster
is reasonably well sampled.  Each scan was repeated several times, and
scans in background fields well offset from the cluster were sometimes
acquired as well.  In the reduction of the scans, very bright isolated
stars were cut out of the part of the image used.
An effort was made to include
several very metal rich galactic globular clusters in this group.

The very metal rich globular clusters are usually seen
against the strong background of the galactic bulge, and proper sky
subtraction, particularly for the most diffuse or faintest of them,
becomes an issue.  We have tried various schemes for sky subtraction,
and can alter the measured $H_{\beta}$ by not more than 10\%, and the
other indices by even less.  NGC 6539 is the most difficult case
among the ones we observed.

The measured indices in the Lick system
for the galactic globular clusters from our Keck+LRIS data are 
listed in Table 3.  We supplement these data by the observations
of Burstein et al (1984).  Six of our sample of galactic GCs 
overlap their sample, and the intercomparison of the indices is
very encouraging.

%
%

We adopt the abundances of Zinn (1985) for the galactic
globular clusters.  
The $UBVRI$ colors for the galactic globular clusters are taken from 
the compilation of Harris (1996).
Several of the galactic globular
clusters lack photometry at $R$ and their $(U-R)$ colors were 
obtained by interpolating for $(V-R)$ from mean relationships
for the galactic globulars given by Reed et al (1988) combined with
their observed $UBV$ colors.  The extinction ratios of Reed et al (1988)
are adopted, specifically $E(U-R)/E(B-V)$ = 2.54.  There
are two galactic GCs in our sample with $E(B-V) > 0.8$, NGC 6539
($E(B-V)$ = 1.00) and NGC 6440 ($E(B-V)$ = 1.09).  In those two
cases, $(U-R)_0$ is quite uncertain. 

It should be noted that these scans cover only the cores/central regions of
the galactic GCs, while our spectra of the M87 GCs include
essentially all the integrated light of each cluster.  If there is strong
mass segregation in galactic GCs, differential shifts in the colors
and indices of the galactic versus the M87 GCs could be produced.
This does not appear to be the case.

\section{QUALITATIVE ANALYSIS}

We present first an analysis which is free of any models (and of the
uncertainty of any models).  Hence it is qualitative in nature,
comparing the M87 GCs with the galactic GCs.
We begun by plotting various indices as a function of $(U-R)$ color, 
adopting the optical photometry of Strom et al (1981) for the M87 GCs.
Suitable photometry is not given in Strom et al (1981) for some of the
clusters in his sample, and is also not available
for the objects close to the
nucleus of M87 that were added in Paper~I.  In some cases we measured
$(U-R)$ (see Appendix B), but for four clusters close to the M87 nucleus no
photometry is available and they are omitted from
the figures where $(U-R)$ is used (Figures 2 -- 8, 18, and 19).
These four M87 GCs are not uniformly at the extreme in any parameter,
and omitting them does not noticeably modify any of the trends displayed
in these figures.


Each of the figures in this section has two panels.  In the first, the 
indices of each of the M87 GCs
are shown.  Large filled circles represent objects with $QSNR \ge 40$.
The 1$\sigma$ 
error bars shown are those appropriate to an object with the minimum
$QSNR$ to be included in the sample.  In the second panel of each figure 
the galactic globular clusters from our sample are shown as ``*'', 
while those from Burstein et al (1984) are shown as open circles. 
The measurements for those clusters 
in common in the two galactic GC samples are connected by
straight lines.  Here
the M87 data is represented as a solid line, the ``median line''
defined by Press et al (1986),
which minimize the total absolute deviation of the points.
The dashed lines are lines with the same slope as the median
line but which enclose 75\% of the points on each side of the median line.
Thus the two dashed lines enclose 75\% of the distribution of the M87 GCs.

%
%

Inspection of the scans of the galactic GCs shows that the most metal
rich among them have about 15\% absorption in the 6250\AA~band of TiO.
The absorption in the 5950\AA~band of TiO which is measured by
the TiO$_1$ index of the Lick system, while still
visible in the scans, is about 1/3 as large.  
We thus expect the TiO$_1$ index to range from 0 in the more metal 
poor galactic (and M87) GCs to about 5\% absorption in the most 
metal rich ones.  

This behavior is in fact shown in figure 2 which
presents the TiO$_1$ index as a function of $(U-R)$ color.  We
use figure 2 to verify that any difference in
global continuum slope between our measurements and
the original Lick IDS spectra is not introducing noticeable errors.
The TiO$_1$ index encompasses a range of 285\AA~(Table 1),
far more than any other index we use.  
Note that this index is expressed as
a magnitude difference, unlike all the others we use, which are
expressed as equivalent widths (EW in Table 1) of absorption.

The TiO$_1$ index becomes more positive as the object
becomes redder, and the total range over the full range
of the dataset is that predicted above.  More importantly, the comparison between the Burstein
et al (1984) indices from Lick observations of galactic globular 
clusters and our indices indicates a
difference of not more than 2\% in this index, with perhaps
a small mean difference in slope.  Given that this
is the broadest index, we feel that continuum slope differences
should not be a serious problem for the other indices we measure.
However a 2\% error in continuum placement
for an index 30\AA~wide corresponds to a 0.6\AA~error in equivalent
width, and the range in strength of absorption from
very metal poor to very metal rich in the
Mg $b$ index for example is only 6\AA, so continuum errors must be 
watched carefully.

Figure 3 shows Mg $b$ versus the color $(U-R)$.  As expected Mg $b$ becomes
stronger in the redder M87 GCs, and the scatter is consistent with
there being a single monotonic relationship between them.  The lower
panel shows that the ``median line'' followed by the M87 GCs is
also followed by the galactic globular clusters.
Figure 4 shows the 5270\AA~Fe blend index versus $(U-R)$, and
Figure 5 shows the 5335\AA~Fe blend index versus $(U-R)$. 
While the total range in these two indices for the M87 GC sample
is not as large as that of Mg $b$, in both
cases again the index becomes stronger as the color becomes redder.
In both cases the galactic globular clusters follow the ``median line''
of the M87 GCs quite closely.

The NaD line index as a function of $(U-R)$ color is shown in Figure 6.
There are six M87 GCs which have extremely strong NaD lines 
(EW$\,>\,$4.5~\AA).  One of them lacks a color measurement
and another, Strom 651, is discussed below; the rest are very
red in $(U-R)$.
Also, the galactic clusters are in the mean displaced
somewhat above the ``median line'' of the M87 GCs.  The object
which has the largest vertical displacement is NGC 6712.  In this case
it is clear that the measured NaD index is being influenced by 
strong interstellar lines.
At galactic coordinates of $(25^{\circ}, -4^{\circ})$,
and with a distance of $\approx$7 kpc,  there is ample opportunity
for substantial interstellar absorption to occur along the line
of sight to this moderately metal poor galactic globular cluster.
The very metal rich galactic GCs also lie above the ``median line'',
implying strong interstellar
NaD absorption.  For objects of reddening $E(B-V) \approx 1$ mag 
in the galactic plane (which is characteristic of the most heavily
reddened of the galactic GCs),
one may expect interstellar absorption in the NaD lines of 
$\approx$1.5\AA~(Hobbs 1974, Cohen 1975).

The M87 GC Strom 651 is also identified on this figure.  A
comparison of the NaD index versus Mg $b$ index plot (Figure 9) 
with Figure 6 
suggests that this M87 GC has either been misidentified by us 
or has an incorrect $(U-R)$ color in Table 3 of Strom et al (1981).
We have checked, beginning with this table, the identification
of Strom 651 on our charts, and have checked that the correct object
was in fact observed by superposing the images of the mask and of the
field in M87 taken to align the slitmask.  The $R$ mag of the object
we observed (see Appendix B) is in good
agreement with that given by Strom et al (1981).

The Balmer line indices $H_{\beta}$ and $H_{\alpha}$ are shown
in Figure 7 and Figure 8 as a function of $(U-R)$ color.  Since
the $H_{\alpha}$ index was defined by us, there 
are no published measurements of $H_{\alpha}$ in the Lick system,
so the number of galactic globular clusters with measured indices is small.  
The index we have defined for $H_{\alpha}$
shows similar behavior to that of $H_{\beta}$,
with the Balmer lines increasing in strength for the
bluer M87 GCs, the opposite of the behavior of the metal
line indices.  The SNR of the spectra is somewhat higher 
at $H_{\alpha}$ than at $H_{\beta}$ and this is reflected in the
relative size of the dispersions in the two plots.

We now display a few index--index plots, avoiding the use of $(U-R)$ colors.
This means that all 150 of the M87 GCs can now be displayed.
Figure 9 shows the NaD line index versus the Mg $b$ index.
Again the galactic globulars are displaced upward from the ``median line''
of the M87 GCs.  NGC 6712, where we are certain that interstellar
NaD lines are contributing to the apparent strength of the NaD index,
has the largest displacement.  Strom 651 now fits onto the main
M87 GC relationship, whereas it was quite anomalous in Figures 3 and 6,
suggesting an error in its $(U-R)$ color.

Figure 10 shows the $H_{\beta}$ index versus Mg $b$ index for the M87 GCs.
As expected, stronger Mg lines are correlated with weaker Balmer lines.
The 5270\AA~Fe blend index is shown as a function of the Mg $b$ index
for the M87 and galactic GCs in Figure 11.

With the exception of the NaD index, where interstellar lines
are believed to contaminate the galactic GC spectra, there is
better agreement between our galactic globular sample and the
Burstein et al (1984) sample in Figures 3, 4, 5, 7, 8, 10 and 11
than occurred for the TiO$_1$ index of Figure 2.  This is also
true of agreement
between the M87 GC ``median line'' and the galactic globulars in
the same set of figures.
We consider this
an indication that continuum slope problems do not affect any
index at a level exceeding 2\%, and perhaps affect the index with
the broadest wavelength coverage, TiO$_1$, at the 2\% level.

\subsection{Qualitative Results}

The following qualitative results emerge from studying these figures:

a) The metallicity range spanned by the M87 GC system is similar 
to that of the galactic globulars, but the M87 GCs have a metal rich 
tail that appears
to extend to significantly higher metallicities than one
finds among the galactic GCs.   
The low metallicity end appears to be similar in the two
systems, but it must be recalled that the Mg $b$ index (and the absorption
in the Mg triplet) are heading rapidly towards zero, and all
discrimination will be lost at that point.  It does appear, however, that
in the M87 GCs the low end of the Mg $b$ distribution
is cut off at a finite Mg $b$ index, and hence loss of
discrimination is not serious.

b) The behavior of the NaD index in the galactic GCs as compared
to the M87 sample shows evidence for a contribution to the galactic GC
NaD indices by interstellar lines.
Since there is not a large
quantity of 10,000K interstellar gas in the halo of M87,
we do not expect substantial interstellar NaD lines 
to occur in the spectra of the M87 GCs. 
 
One might wonder if the
few (six) M87 GCs with very large NaD indices also suffer interstellar
absorption from gas within M87.  These objects all have $R \le 200$ arc-sec
from the center of M87, but are not confined closely to the nucleus
of M87, where one might expect any
gas to be concentrated.  Furthermore they are among the most
metal rich objects in {\it all} 
indices, not just in the NaD index.  We hypothesize that
the behavior of the NaD index just happens
to be more extreme (i.e. more nonlinear) at very high metallicities
than the 
other indices.  That may
be a function of the saturation of the lines, the detailed behavior
of the curve of growth, etc.

c) The relative abundances of Fe, Na, and Mg appear to be changing
together, in the same way as in the galactic globulars
and in galactic stars.  There
are essentially no objects with extremely strong Fe lines but 
weak Mg lines, etc.  In other words, to first order
the nucleosynthesis history appears to have been the same in the
M87 GC system as in the Milky Way GC system.

d) The Balmer line indices also behave similarly in the M87 and in
the galactic globular clusters.  It is well known that the Balmer
lines are the most readily detectable age sensitive features
in the integrated light spectra of old stellar systems
(Rabin 1982, Worthey 1994).  This similarity implies that the mean
age and the age distribution are similar in these two systems.

\subsection {Radial Gradients and Trends with Luminosity} 

Before proceeding to a more quantitative treatment, we address
several other issues in a qualitative way.  

Possible radial gradients are explored in Figure 12,
which displays three of the indices as a function of projected radius $R$. 
The large scatter at all radii in the metallicity of the GCs in M87 makes
the detection of a radial trend difficult. There is
no trend in the $H_{\beta}$ index with $R$, which 
effectively translates into no trend of GC age
with $R$.  There is a small radial trend in 
Mg $b$ index, which
corresponds to a radial gradient of $\approx$0.2 dex 
over the range $30 \le R \le 500$ arc-sec.
The radial gradient inferred from
the NaD index is about twice as large, but the apparent size
of the radial gradient in the NaD index
is influenced by the six M87 GCs with extremely high NaD
indices.  The gradient for a least squares fit is in this case
significantly larger than for the ``median line'' fit.
However, at all values of $R$ the scatter is much larger than any gradient.
Since the observed radial gradients in the line indices
are small, no effort was made to deproject them.

Although our sample of M87 GCs was selected from the 
magnitude limited photometric survey of Strom et al (1981),
there are a number of effects that lead to the mean brightness
of the objects in our sample 
becoming progressively fainter at larger $R$.
The most important of these is the rapid decrease in the areal
density of M87 GCs as $R$ increases.  This means in practice that
in designing slit masks, one will run out of bright objects
in the outer part of the M87 halo long before one does so in
the inner part of the halo.  A second, less important, factor
has to do with how the sample was defined.
Our selection criterion is based not on brightness, but
on $QSNR$, which in turn depends also on the brightness of the background.
Hence a GC must be brighter to make it into the sample when the
background from the M87 halo becomes brighter, i.e. at smaller
$R$.

We have established that there is a radial gradient in metallicity in
the M87 GC system of the expected sign (more metal rich at smaller $R$), 
and hence this, coupled with a correlation of mean brightness with
projected radius,
will introduce a spurious trend of increasing metallicity 
with increasing brightness.  This is in fact seen in our data.  
Although it is possible that the most massive GCs may have been able
to self-enrich to some extent, increasing their mean metallicity,
no analysis of a magnitude limited photometric sample has ever detected
any correlation of mean color with brightness.
Thus, we assume this is a spurious result introduced
by the biases in our sample selection.

The objects with high $QSNR$ are among the brighter M87 GCs, as we 
would expect.

\subsection{The Histogram  of Mg $b$ Indices for the M87 Globular 
Cluster System}

The histogram of Mg $b$ indices (Figure 13) has a broad peak with a 
metal rich tail.  While there is in principal
little discrimination at extremely low metallicities,
there are few very metal poor clusters in the M87 GC system. 
The 1$\sigma$ error in the Mg $b$ index is 
0.5\AA~for a M87 GC with $QSNR$ at the
minimum required to get into the abundance sample.  
(The binning used in Figure 13 is 0.25\AA/bin.)

The metallicity distribution of the galactic GCs is bimodal, with
peaks at [Fe/H] of $-$1.6 and $-$0.6, using the data of Zinn (1985).
The overall mean [Fe/H] is $-$1.3, and the rms dispersion is
$\sigma=0.6$ dex, which translates into $\approx$1.4\AA~in the 
Mg $b$ index.  The main metal poor peak at [Fe/H]$\,=\,$ $-$1.6 in 
the metallicity histogram of galactic GCs
is approximately Gaussian with a dispersion of about 0.3 dex.

The Mg $b$ histogram for the M87 GC system has a FWHM
comparable to that of the galactic GC system, 
but the mean Mg $b$ index corresponds to
a much higher metallicity.  
The mean value of Mg $b$ in the M87 GCs is 2.28 \AA~(with a formal 
value for $\sigma$ of 1.16 \AA) and the median
value is 2.18 \AA.  
Based on Table 3, these correspond to [Fe/H] $\approx -0.85$ dex
with $\sigma \approx 0.5$ dex.
A comparison with previous determinations of these parameters will
be given in Section~7. 

It is clear that a single Gaussian model cannot fit 
the form of the histogram in Figure~13.
To quantify this observation, 
we have applied a ``mixture-modeling'' algorithm known as 
the \kmm\ test.
The \kmm\ test is an implementation of the ``Expectation Maximization''
algorithm of Dempster et al.\ (1977) and is
described in detail by McLachlan \& Basford (1988).
Use of the \kmm\ test for estimating the significance of
multiple (Gaussian) populations in univariate astronomical datasets
has been explored by Ashman, Bird, \& Zepf (1994).
The test does not involve binning of the data.
Because of the great interest in bimodal metallicity
distributions in giant ellipticals (e.g., Ashman \& Zepf 1992;
Whitmore et al.\ 1995; Geisler et al.\ 1996)
and in order to limit the number of free parameters,
we confine the tests to 
double versus single Gaussian models.

One must take care in performing and interpreting such multiple
component fits.
For instance, blind application of the \kmm\ test to the entire Mg~$b$ 
distribution in Figure~13 finds that the double Gaussian model is to
be preferred with a very high significance, with the peaks of the
Gaussians located at 2.0 and 4.3~\AA, i.e., it uses the second Gaussian
to fit the high-Mg~$b$ tail, containing only 14\% of the GCs.

We wish to see whether the double Gaussian model is preferred for
the main body of the histogram, however.  If the tail is removed by
excluding GCs with Mg~$b > 4.0$~\AA,
then the \kmm\ test prefers the double Gaussian with a 
marginally significant probability of 87\%.  The Gaussians in this
case are centered at 1.46 and 2.73~\AA, containing respectively
55\% and 45\% of the total number of GCs in this Mg~$b$ range.
The common dispersion of the Gaussians is $\sigma{\,=\,}0.60$~\AA.
If instead the clipping is done for Mg~$b > 3.5$~\AA,
the significance level shoots up to 97\%, and the peaks are 
at 1.29 and 2.50~\AA, containing 48\% and 52\% of the GCs,
respectively.  The common dispersion here is $\sigma{\,=\,}0.50$~\AA.
Thus, in neither of these cases would the two components be clearly
distinguished in the histogram, as the separation between the 
Gaussians is just over 2$\sigma$.
The high significance level returned by \kmm\ in the latter case
is likely due to the artificial boxiness caused by the overly
zealous clipping (see the discussion by Ashman et al.\ 1994).
We conclude that the actual significance level of the bimodality,
with the tail excluded, is 85--90\%.
The following section revisits the bimodality issue in the
context of the final [Fe/H] distribution.


\section{SEPARATION OF AGE AND ABUNDANCE USING WORTHEY'S MODELS}

\subsection{Calibration of the Worthey Models}

We now proceed to a more quantitative treatment of the data.
We use the models of Worthey (1994), assuming a single burst of
star formation and a Salpeter IMF.  The choice of IMF is
not critical as long as the integrated light is dominated by
the giants rather than by the main sequence dwarfs.
It is sufficiently difficult just to extract an abundance
and an age for each M87 GC that we chose to avoid the 
complications of the additional
parameters introduced by the models of Borges et al (1995).

We will use the Worthey (1994) models to assign definite metallicities
and ages to the M87 GCs.  First we need to rescale the
predictions of these models to reproduce the metallicities
we have adopted for the galactic GCs, namely those of Zinn (1985).
We define a canonical metallicity
(henceforth [Fe/H]$_W$) as the metallicity predicted 
by the Worthey (1994) model
for a fixed age of 12 billion
years using a $\chi^2$ minimization to the Mg $b$, NaD, Fe5270 and 
Fe5335 indices together with their appropriate errors.  
The $H_{\beta}$ index is ignored here.  Tests using 
simultaneous fitting for both age and metallicity with the Worthey (1994)
model grid show that [Fe/H]$_W$ differs little from that of a two
parameter fit.

Next we rescale the [Fe/H]$_W$ values inferred for the 
galactic GCs using the
data in Table 3 with the Worthey (1994) models to produce
agreement with the Zinn (1985) metallicities for the galactic GCs.
This rescaled value is denoted as [Fe/H]$_Z$.
Figure 14 shows [Fe/H]$_Z$ values from Zinn (1985) versus
[Fe/H]$_W$ from the Worthey models for the galactic globular
clusters and indicates the scaling required.
A linear transformation, 
[Fe/H]$_Z$ $= 0.760 {\times}$ [Fe/H]$_W - 0.265$, is satisfactory.
The new data for the galactic GCs 
that were actually used to derive the fit
are shown as filled circles, and open circles denote
the data of Burstein et al (1984).  The symbols representing clusters 
in common between the two samples are connected by lines.
%
%

The agreement between the predictions of the Worthey models and  
the data for the M87 and galactic GCs is shown in Figures
15 through 19.  These figures display only the 
behavior of the M87 GCs, but we have
already demonstrated in figures 2 through 11
that the behavior of the galactic globular
clusters follows the ``median line'' of the M87 GCs.
Figure 15, 16, and 17 are index--index plots 
(NaD versus Mg $b$, 5270 versus Mg $b$, and $H_{\beta}$ versus Mg $b$)
onto which the predictions of the Worthey (1994) models have been
overlaid. Figure 18 displays Mg $b$ versus the $(U-R)$ color with the
predictions of the Worthey models,
while Figure 19 does the same for the $H_{\beta}$ index.
The Worthey (1994) models range from $3 \times$ solar to 1/100 solar 
metallicity (which values have to be rescaled to our adopted Zinn scale)
with ages of 2, 3, 5, 8, 12, and 17 Gyr.  The metal poor
models were only computed for ages of 8, 12, and 17 Gyr,
but that does not matter, as no M87 GC turns out to be younger
than 6 Gyr.
The sequence of models of a fixed metallicity are connected by a line,
and the prediction for each age is given by a ``x'' along the line.

One has the overall impression from looking at these figures that the
Worthey (1994) models are a reasonable representation of the data.
Note the very different behavior predicted for $H_{\beta}$ as
a function of age (and metallicity) than for the NaD and Mg $b$ indices,
as supported by the data for both the M87 and the galactic GCs.
The ability of the Worthey models 
to reproduce the mean behavior of the $H_{\beta}$ index
at all values of the Mg $b$ index implies that the assumptions made
regarding the contribution of the horizontal branch (HB) and of
more advanced stages of stellar evolution to the total integrated light
in these models are basically
correct.  These assumptions include that the standard HB for metal rich
clusters is a red clump and there is no extended blue HB such as was
recently found in several metal rich galactic GCs by Rich et al (1997).
Over the wavelength range considered here, if such an extended blue HB
is present in metal rich GCs, its contribution to the integrated light
must be small.

The very strong NaD lines seen in six of the M87 GCs
are partially reproduced by the most
metal rich of the Worthey models, but not completely.

Figure 16 shows the most glaring failure of the Worthey (1994) models,
in that the 5270\AA~Fe indices of the M87 GCs are always smaller
than predicted by the models at the metal rich (high Mg $b$) end.
This is reminiscent of the problems encountered by Worthey et al (1992)
in trying to match the indices of elliptical galaxies.

We thus have some, although not perfect, confidence in the validity of the Worthey (1994) models
for the integrated light of stellar systems,
and hence in the ages and metallicities derived by
applying them to our data on
the M87 GCs.

The Worthey (1994) models were extrapolated to 1/200 solar metallicity.  
Taking into account the rescaling of the Worthey abundance scale, the
minimum possible metallicity returned by the fit is [Fe/H]$_Z = -2.2$ dex,
while 1.3 $\times$ solar is the
maximum.   Also the metal poor models were extrapolated to an age of 5 Gyr,
which is the minimum possible age for a metal poor model.
The age range returned by the fit is 5 to 17 Gyr.

Figure 20 shows the metallicity distribution
([Fe/H]$_Z$) for the sample of 150 M87 GCs.
The mean [Fe/H]$_Z$ is $-0.98$ dex, with
$\sigma = 0.5$ dex, while the median [Fe/H ]$_Z$ is $-0.95$ dex. 
The rms error in [Fe/H] due only to measurement errors for the Mg $b$
index 
is 0.22 dex for an object with $QSNR$ such that it just gets into the
abundance sample.  This abundance error is actually slightly degraded by adding
in the Na and Fe indices because of their smaller range (with the
same measurement error).
Overlaid on this metallicity distribution
is the histogram for the galactic GCs represented by a solid line
with data from the online compilation of Harris (1996) which is on the 
same metallicity scale (that 
of Zinn 1985) that we have adopted here.
Just as in Figure 13 the M87 GC system shows a broad peak with a 
possible hint of multiple components.  An
artificial pileup of five objects at
each of the extremes of the allowed range for metallicity, 
$-2.2$ and +0.1 dex, is seen in Figure 20, 
while in Figure 13 these objects are shown more correctly as tails in the  
distribution of the Mg $b$ indices.  It is clear from figures
3, 9, and 13 that the objects
at the upper limit of [Fe/H] returned by our fit
are in fact more metal rich than any
of the galactic GCs.

We performed the \kmm\ test on the final list of [Fe/H]$_Z$ values,
excluding those artificially piled up at the extremes.  
The test results indicate that the double Gaussian model is preferred
over the single one at the 89\% significance level.
The two ``subpopulations'' in this model are
located at $-$1.3 and $-$0.7 dex and contain 40\% and 60\% of the total,
respectively.  The common dispersion is $\sigma{\,=\,}$0.3~dex.
These results are similar to those found for the Mg~$b$ distribution,
namely that there is marginal evidence, at the $\lesssim$90\% level,
for bimodality in the metallicity distribution.
The locations of the
metallicity ``peaks'' found here correspond well with those 
inferred from the Mg~$b$ \kmm\ test results.

For the halo of M87, the Dressler et al (1987) measurement of the
Lick indices can be transformed into [Fe/H] using the Worthey models.
Angeletti \& Giannone (1997) do this, extrapolating
the measurements made on the nucleus to a projected
radius of 200 arc-sec, and find [Fe/H]$_Z = -0.3$ dex, where
we have transformed their value to our adopted metallicity scale.
This is about a factor of 4 higher than the metallicity of the
M87 GCs at a similar projected radius.  All photometric studies of
the M87 GC system, beginning
with the work of Strom et al (1981), have found that the mean color
of the GCs at a particular $R$ is significantly bluer than the light
of the halo of M87 itself at that projected radius.  This color
difference is produced by the difference in mean metallicity discussed
above.

\subsection{Determination of the Age Distribution of the M87 GCs}

We intend to apply the Worthey (1994) models as described above to
predict both the age and metallicity of a M87 GC from the full set of
indices, Mg $b$, NaD, Fe 5270, Fe 5335, and $H_{\beta}$.
Since the age discrimination is coming primarily from the $H_{\beta}$
index, the age is extremely sensitive to errors in that index.
We therefore calculate the ages using the ``median lines'' for
$H_{\beta}$ rather than the values of the $H_{\beta}$ index for
each individual object.  In effect, we are calculating the median
age as a function of metallicity, since the $H_{\beta}$ lines are
determined with respect to the metal indices.

First we test the scheme using the combined sample of galactic globular
clusters, our data plus that of Burstein et al (1984).  The histogram
of ages for this sample of 23 galactic GCs is given in Figure 21.
The cross hatched area denotes our sample of galactic GCs, while
the open fill denotes the merged sample. (For all the galactic
GCs with $E(B-V) > 0.4$ mag, the NaD index
was not used, as it is heavily contaminated by interstellar absorption.)
A number of galactic GCs piled up at the upper limit of
the fit, 17 Gyr.  Figures 10, 17 and 19 demonstrate the problem, 
namely the
metal rich galactic GCs tend to lie below the ``median line'' determined
by the much larger M87 sample, i.e. their $H_{\beta}$ indices appear
weaker than those predicted even by Worthey's (1994) oldest models,
while the metal poor ones tend to
lie above it, producing the two peaks in the histogram of age shown
in figure 21.  Thus, at least formally, we find the metal rich
galactic GCs to be older than the metal poor ones.

The median age of the galactic GCs is 15 Gyr. 
While no special credence should be placed
in this value, it is well within the range of ages normally ascribed
to the galactic GCs, and hence we believe that this procedure will give
an accurate determination of the relative age of the M87 and galactic
GC systems.

We now apply the Worthey (1994) models to our M87 data.  We averaged
the indices from the two weaker Fe blends 5270 and 5335, using the
average Fe plus NaD and Mg $b$ indices (with their appropriate errors) 
and the $H_{\beta}$ ``median lines''.
The result of this process is the age distribution for the M87 GC
system shown in Figure 22.  The pileup at 17 Gyr is
artificial and is
due to the maximum age returned by the fitting process.
The mean age is 13.2 Gyr with a 1$\sigma$ deviation of 2.2 Gyr and
the median age is 13 Gyr.  This is identical to within the errors
to the age of the galactic GC system obtained above.

We next assess the impact of observational errors in producing the
spread seen in Figure 22.  The 1$\sigma$ error in the 
$H_{\beta}$ index for a M87 GC with
the minimum $QSNR$ to get into our abundance sample was given
in Section 3.3 as 0.5\AA, while the rms deviation of the 148
points around the ``median line'' for the $H_{\beta}$ index is
0.55\AA.  (The spectra of two of the M87 GCs do not reach
blue enough to allow measurement of the $H_{\beta}$ index.)
%
%
%
Increasing the
$H_{\beta}$ index by 0.5\AA~reduces
the derived age from 13 Gyr to 6 Gyr, while decreasing that index
by the same amount increases the age from 13 Gyr to the upper limit 
of the fits at 17 Gyr.  Given our sample size, the uncertainty in the
``median line'' at any point is $\approx$0.045\AA.  A change
of this size in the $H_{\beta}$ index changes the deduced age
by $\approx$0.9 Gyr.
Thus while observational errors seriously impact the age of an
individual GC inferred from its measured
$H_{\beta}$ index, use of the ``median line'' for the
$H_{\beta}$ index should produce 
a good estimate of the age.

There is no evidence for a substantial population of M87 GCs
younger than 12 Gyr.  The median age and the age distribution of the M87 GCs
and of the galactic GCs are identical.

Table 4 gives the age estimates for the M87 GCs divided into
six metallicity bins, each containing 25 objects.  There is no 
sign of a variation in age with metallicity.  One should note, however,
that the galactic globular clusters emerge from the above
analysis with an age distribution such
that formally the metal rich galactic GCs are older than the metal poor
ones.  We do not understand the cause of this.  If it arises from an error in
Worthey's (1994) models, then correcting this error might make
the metal rich M87 GCs somewhat younger than the metal poor ones.

\subsection{Correlations of Velocity Dispersion with Metallicity}

We have looked for a correlation between velocity dispersion and metallicity
among the M87 GCs in our sample, combining the abundances given here
with the $v_r$ of paper 1.  Such a correlation is
predicted by various theories of globular cluster formation, particularly
if one chooses to believe that the metal rich population of GCs is
a ``disk population'', while the metal poor GCs represent the halo.
In this case one would expect the metal poor GCs to show a larger
velocity dispersion, while the metal rich ones, with their
smaller velocity dispersion, might show
rotation, as is seen in the Galaxy (Zinn 1985).  

Subdividing the sample of 150 M87 GCs
into halves, and then into thirds, does not
show the expected effect. If anything, the velocity dispersion of
the metal poor GCs is smaller than that of the metal rich ones.

This issue is complicated by exactly what one adopts for the
rotation curve for the stellar component of M87 and for the M87 GCs
(see Paper I).

\section{COMPARISON WITH PREVIOUS STUDIES AND WITH STUDIES OF NGC 1399}

\subsection{Comparison with Previous Studies of M87}

As discussed earlier there have been many photometric studies of the
M87 GC system, both from the ground and from HST, as well as initial
spectroscopic studies of small samples.  The mean metallicity
we have derived for the M87 GC system, $-0.98$ dex, with
$\sigma = 0.5$ dex on the scale of Zinn (1985) is in good agreement
with the earlier estimates summarized in Table 5.

%
%

The beautiful X-ray data from Ginga (Koyama et al 1991) and from
ASCA (Matsumoto et al 1996) extend from the nucleus to $R = 40$ arc-min 
and show definite small radial gradients in O, Si, S, and Fe, with
the gradient in O being the weakest.  The abundances they derive
for the hot X-ray emitting gas are significantly higher than
those inferred from the M87 GC system and are comparable to those
of the mean of the M87 stellar halo.
This might be due to the
buildup of heavy elements in the gas since the initial formation
of the M87 GCs through mass loss from the M87 halo stars.

The subject of radial gradients in the M87 GC system (see the review
by Harris 1991) has been quite controversial.  Although Strom et al (1981)
claimed to detect a radial gradient in the mean broad band colors
of the M87 GCs, this was not reproduced by subsequent, presumably more
accurate, CCD based studies by Cohen (1988) and by Couture et al (1990).
However, Lee \& Geisler (1993) using Washington photometry
claimed to detect a radial gradient of a factor of 4 in mean abundance
over the range in projected radius of their observations, 60 to 500 arc-sec.

It is clear from our work that there is a radial gradient in 
mean abundance in the M87 GC system, but that
the dispersion in metallicity at all radii
is so large that its easy to lose in the
observational errors.  It is a tribute to the metallicity sensitivity
of the ultraviolet colors used in the
original photographic survey of Strom et al (1981) and of the
Washington narrow band photometric system used by Lee \& Geisler (1993),
that they detected this gradient. The
large size of these two samples and their large range in radial coverage
also contributed to detection of these gradients.  The calibration
of photometric indices versus metallicity for these two
photometric systems is derived using our [Fe/H]$_Z$ values for the
M87 GCs together with published photometry for these objects in appendix C.

The exact size of the radial gradient of abundance within the M87 GC
system is still not well determined, but the existence of a gradient
is now beyond doubt.

Meanwhile the high precision HST photometric surveys of the M87 GC system by
Whitmore et al (1995) and by Elson \& Santiago (1996a,b) have as their major
new result the claim that the abundance distribution of the M87
GCs is bimodal, with peaks which agree between these two studies
at $(V-I) \approx 0.95$ and $\approx$1.20 mag.  Using the 
transformation of Couture et al (1990), which is on the Zinn (1985)
abundance scale, these colors
correspond to [Fe/H] = $-$1.3 and 0.0 dex.  They do not agree well with
the histogram of abundances for the M87 GC system we obtained shown
in Figure 20.  Use of the transformation of Kissler-Patig et al (1997),
which was developed for very red objects, suggests the upper peak
has [Fe/H] $-0.5$ dex, much closer to what we see in our M87
GC metallicity distribution.  The radial gradients in the M87 GC
system also need to be taken into account here, as the
HST studies are heavily biased towards the center of M87 due to the
small spatial field of WFPC, while our analysis covers a much larger
range in projected radius.

\subsection{Comparison with Results for NGC 1399}

The only other giant elliptical whose GC system has been studied in
some detail is NGC 1399, the cD galaxy in the Fornax cluster.  There
have been many broad band photometric studies (see, for example,
Bridges et al (1991) and a study using the Washington system
by Ostrov et al (1993) which augments an earlier effort by
Geisler \& Forte (1990).  Very recently Kissler-Patig et al (1997)
presented an initial spectroscopic survey of a small sample of GCs
in NGC 1399 using LRIS at Keck.  The available results regarding
the abundances in the NGC 1399 GC system are quite similar to
what we find in the M87 GC system.  The mean [Fe/H]
for the NGC 1399 GC system is $-0.90 \pm0.2$ dex,
with a spread ranging from solar to very metal poor.  Ostrov et al (1993)
report a small radial gradient.  

\section{IMPLICATIONS FOR GALAXY AND GLOBULAR CLUSTER FORMATION}

Harris (1991) reviewed the many theories that have been advanced
to explain the formation of globular clusters and their confrontation
with the mounting body of known facts about these systems.
In recent years, more advanced numerical simulations of galaxy mergers
(e.g., Barnes \& Hernquist 1992)
and isophotal evidence such as shell structures (e.g., Quinn 1984)
have convinced many people that mergers of spiral galaxies
may play an important role in the formation and evolution
of elliptical galaxies.
Ashman \& Zepf (1992, 1997) have emphasized the possible role of mergers
on the formation of GC systems and have claimed that they can 
reproduce the high $S_N$ values
(a parameter that characterizes the number of globular
clusters per unit halo luminosity) seen in systems such as M87
under such circumstances.

Our work presents another set of constraints for these models.
The constraints we have placed on the radial velocity
distribution (see Paper~I), the metallicity distribution,
the radial abundance gradients, and the age distribution 
within the M87 GC system
suggest that mergers, if important, all happened near the beginning
of the collapse of the Virgo galaxy cluster.  Everything is so well
organized, and so continuous with the properties of less
luminous ellipticals, that mergers of fully developed galaxies
to form ellipticals seems quite unlikely.  Global correlations such
as that between the metallicity of the GC system of an elliptical galaxy
and its luminosity, first pointed out by van den Bergh (1975),
are now even stronger.  

Our major new contribution to this process is that   
we have established the timescale for
the process of the assembling of the GC system of
a galaxy like M87; it seems to have been less than 2 Gyr.  This
is a time span comparable to the age range believed to exist
among the galactic GCs (see VandenBerg et al 1996).

Theories where the GC system of the central cD galaxy in a cluster
of galaxies is formed by accreting GCs from neighboring galaxies
through orbital effects,
such as that advocated by Muzzio (1987) can be ruled out on the
basis of the extended timescale for GC formation.  Formation
of GCs in the cooling flow around the central cD
in clusters of galaxies advanced by Fabian et al (1984)
can also be ruled out for similar reasons.

Another important point to note where our work establishes a new constraint
is that the relative abundances of Fe, Na, and Mg appear to be changing
together in the M87 GC system
in the same way as in the galactic globulars.
In other words, to first order
the nucleosynthesis history appears to have been the same in these
two GC systems.  The same is true for the M31 GC system (Cohen et al 1997).  
The details of chemical evolution of
GCs depend on  many complex and poorly understood phenomena.  These include
the process of galaxy formation, the halo formation process,
star formation within the proto-globular cluster,
the formation of supernovae, the nuclear burning within supernovae
and the dispersal of their ejecta, the processes of stellar evolution,
etc.  Somehow all of this happened so as to 
produce very similar and very reproducible
trends in the oldest objects we can study.  These include 
very distant objects at high redshift in the young universe
that we perceive through the QSO absorption lines that they produce
(Lu et al 1996)
as well as old objects in the local universe, such
as the M87 GC system, the galactic GC system, or the galactic halo.

It is very reassuring that in some sense the laws of physics
work over such enormous ranges in distance and time.  Obviously
as one looks in more detail, as is possible within our own galaxy,
one begins to see finer and finer effects.
High precision studies of field halo stars, as well as of stars in 
galactic globular
clusters, have established that in our galaxy there are a number of
prominent changes in abundance ratios, such as the increase of
$\alpha$--element abundances relative to Fe in stars with
[Fe/H] between --2 dex and solar abundance, as reviewed by 
Wheeler, Sneden \& Truran (1989).
This particular trend is believed to be due to the time delay involved for
Type Ia supernovae to become effective sources of heavy elements.
These trends have been further extended to even lower metallicity
by several groups, most
recently McWilliam et al (1995). 
But the gross trends appear to be universal.  We can now 
examine in considerable detail the brightest
early type supergiants in the nearest Local Group
galaxies, (McCarthy et al 1995, Monteverde et al 1997), to find
the same gross trends repeated.

\section{SUMMARY}

In Paper~I we used radial velocity data to isolate a sample of
205 bona fide members of the M87 globular cluster system.
Here we have selected a subset of this sample for more
detailed analysis on the
basis of the signal-to-noise ratio of their spectra.  This sample
includes 162 spectra of 150 M87 GCs.  

We used the Lick indices to
measure the strength of the stronger absorption line features
because of the extensive database available for stars and the integrated
light of stellar systems.  We included
Mg $b$, 5270 Fe, 5335 Fe, NaD, and TiO$_1$, as well as the
$H_{\beta}$ index, and added an index for $H_{\alpha}$.

First, we qualitatively compared the behavior of the M87 GCs and
the galactic GCs.
We combined new scans of 12 galactic GCs with the data of Burstein et al (1984)
for a total sample of 23 galactic GCs.
A study of the index--index plots and of the index -- broad band $(U-R)$ 
color plots shows that the metallicity range spanned by the M87 GC system
is similar to that of the galactic GCs, but the former has a metal rich
tail extending to significantly higher metallicity than one finds
among the galactic GCs.  

There are 6 M87 GCs with extremely strong NaD indices.
The NaD indices for the galactic GC system
shows definite evidence for the influence of interstellar absorption.

The relative abundances of Fe, Mg, and Na in the M87 GCs follow similar
trends to those of the galactic GCs.  To first order, the nucleosynthesis
history in the M87 GC system appears to have been the same as in the 
Milky Way GC system.  The behavior of the Balmer lines is also quite 
similar, implying a similar mean age and age distribution in the two
systems.

There is a small radial gradient in the metal line indices in the sense of
a decline in metallicity as $R$ (projected radius) increases, but there
is no detectable radial change in the $H_{\beta}$ index.  No dependence
of dispersion of radial velocity with metallicity was detected.

To make the results more quantitative, we used the Worthey (1994) models
for the integrated light of stellar systems assuming a single
burst of star formation and a Salpeter initial mass function.
We adopted the metallicity scale of Zinn (1985) for the galactic GCs.
We checked the validity of the Worthey (1994) models and 
normalized their metallicity scale to the Zinn scale by analyzing
the indices of the galactic GCs.  

The age dependence comes
almost entirely from the Balmer lines. In consideration
of the observational errors, we combined the measured
metal line indices for each object with  the ``median lines'' for the
$H_{\beta}$ index, rather than with the individually measured values for
each GC, to derive the age of each galactic GC.  The result is
a median age of 15 Gyr for the sample of 23 galactic GCs.
 
We then applied this formalism to the M87 GC system.  The median age
for the 150 M87 GCs is 13 Gyr, with $\sigma$ = 2 Gyr.  The median 
metallicity is [Fe/H]$_Z = -0.95$ dex with $\sigma$ = 0.5 dex,
in good agreement with previous work.  There is no {\it obvious}
bimodality in the abundance distribution,
but the peak of the distribution is very broad.
Moreover, application of the \kmm\ test to the abundance distribution
finds marginal evidence for bimodality, with peaks near
[Fe/H]$_Z \sim -1.3$ and $-$0.7.
We compared these results to analyses of the GC system
of NGC 1399, the cD giant elliptical at the center of the Fornax cluster,
to find very similar properties for the two systems.

The major new contributions we have made to this field are
twofold.  The first is the
establishment of the age distribution for the M87 GCs.  The
mean age for the M87 GCs is
comparable to that of the galactic
globular clusters, with a small dispersion ($\sigma$ = 2 Gyr) about
that value.  The second concerns the details of chemical evolution and
enrichment in GC systems.  We have found the same trends prevail
among the abundances of Fe, Mg, and Na as occur in the
galactic GC system.  The former is a very powerful constraint and
we have briefly discussed the implications for theories of globular
cluster formation.

\acknowledgements
The entire Keck/LRIS user community owes a huge
debt to Jerry Nelson, Gerry Smith, Bev Oke, and many other people who
have worked to make the Keck Telescope and LRIS a reality.  We are grateful to
the W. M. Keck Foundation, and particularly its late president, Howard
Keck, for the vision to fund the construction of the W. M. Keck
Observatory.  

JGC is grateful to NSF grant AST96-16729 for support.
JPB is grateful to the Sherman Fairchild Foundation for support.

\newpage

\begin{appendix}

\section{NEW RADIAL VELOCITY MEASUREMENTS}
 
In the process of re-analysis of the spectra of the M87 GCs to
prepare the final abundance sample, five M87 GCs were found for
which the final spectra gave $v_r$ different from the values published
in Paper~I by more than 150 \kms.   In addition, Strom 682, which 
was not included in the $v_r$ sample of Paper~I,
now has a definite $v_r$.  These values are listed in Table A1.

\section{NEW $U$, $R$ PHOTOMETRY}

A few of the M87 GCs  in our sample do not have $UR$ photometry from
Strom et al (1981).  Also the added sample of nuclear GCs in M87 has
no published photometry.  We therefore obtained a set of images in $U$ and in 
$R$ using COSMIC (Dressler 1993) at the prime focus of the 5--m Hale telescope
at Palomar Observatory.  Several COSMIC fields were required
to cover the large angular field on the sky of the M87 GCs.   
Because of  the overlap between fields, most of the M87 GCs for 
which colors are required appear on several of the fields and hence 
have multiple measurements.  The total integration time/field is 100 sec
for $R$ and 1200 sec for $U$.  We used Strom et al's (1981) photographic
photometry to define the zero points for each color.
Table B1 lists our newly measured $U$ and $R$ magnitudes.

We thank Charles Steidel for use of his $U$ filter.

\section{CALIBRATION OF $(U-R)$ AND WASHINGTON PHOTOMETRY AS A FUNCTION OF
METALLICITY}

At the request of the referee, as an aid to future investigators
we have derived a calibration
of the $(U-R)$ photometry of Strom et al (1981) in terms of our
metallicities for the M87 GCs.  A linear least squares fit 
for the 133 M87 GCs in our sample with $(U,R)$ photometry
from Strom et al (1981) or from Table B1 and with  $-2 < $[Fe/H]$_Z < 0$ dex
gives [Fe/H]$_Z$ = $0.81 {\rm{\times}} (U-R)_0 -2.25$ dex,
with a rms residual of 0.29 dex.  Note that the dereddened color
is used here; $E(U-R) = 0.06$ mag for M87.  A linear fit
is adequate.  This dispersion is consistent with the
uncertainties of our abundances coupled with those of the photometry.

We have also done this for the Washington photometry of the M87 GCs by 
taking the [Fe/H] values obtained by Lee \& Geisler (1993)
for a sample of M87 GCs (Table 3 of their paper)
and identifying these objects
in the sample of Strom et al (1981).  Since there are several
inconsistencies in Table 3 between the names of the fields 
in the table and the names of the fields on the charts of this paper, this was
non-trivial.  Because the area observed
by Lee \& Geisler
excludes most of the southern part of M87 and half of the East2 field
(as identified on their charts)
is beyond the area of the Strom et al (1981) sample from which
we picked our M87 GC candidates, there are only 
38 objects in common.  The resulting linear fit is [Fe/H]$_Z$ =
$1.01 {\rm{\times}}$ [Fe/H]$_{LG} + 0.16$ dex with a rms residual of
0.39 dex.  This fit appears to fail at the highest metallicities.
A quadratic fit gives only a slightly smaller dispersion.

\end{appendix}

\newpage

%
\begin{deluxetable}{lcccc}
\tablewidth{0pt}
\scriptsize
\tablecaption{Bandpasses for Absorption Indices}
\tablehead{
   \colhead{Index} & \colhead{Blue continuum (\AA)} 
 & \colhead{Feature bandpass (\AA)}
 & \colhead{Red continuum (\AA)} & \colhead{Type}
}
\startdata
H$\beta$ &  4829.50--4848.25 & 4849.50--4877.00 & 4878.25--4892.00 & EW \\
Mg$_1$   &  4897.00--4958.25 & 5071.00--5134.75 & 5303.00--5366.75 & Mag \\
Mg$_2$   &  4897.00--4958.25 & 5156.00--5197.25 & 5303.00--5366.75 & Mag \\
Mg $b$   &  5144.50--5162.00 & 5162.00--5193.25 & 5193.25--5207.00 & EW \\
Fe5270   &  5235.50--5248.00 & 5248.00--5286.75 & 5288.00--5319.25 & EW \\
Fe5335   &  5307.25--5317.25 & 5314.75--5353.50 & 5356.00--5364.75 & EW \\
NaD      &  5863.00--5876.75 & 5879.25--5910.50 & 5924.50--5949.25 & EW \\
TiO$_1$  &  5819.00--5850.25 & 5939.00--5995.25 & 6041.00--6104.75 & Mag \\
H$\alpha$&  6420.00--6455.00 & 6548.00--6578.00 & 6600.00--6640.00 & EW \\
\enddata
\end{deluxetable}

\begin{deluxetable}{lrrrrrrrrrrc}
\scriptsize
\tablewidth{0pt}
\tablecaption{Index Measurements for M87 Globular Clusters}
\tablehead{
\colhead{ID} & \colhead{H$\beta$} & \colhead{Mg$_1$} & \colhead{Mg$_2$} & 
\colhead{Mg $b$} & \colhead{Fe5270} & \colhead{Fe5335} & \colhead{NaD} &
\colhead{TiO$_1$} & \colhead{H$\alpha$} & \colhead{QSNR} & \colhead{[Fe/H]$_Z$} 
\\
\colhead{} & \colhead{(\AA)} & \colhead{(mag)} & \colhead{(mag)} & 
\colhead{(\AA)} & \colhead{(\AA)} & \colhead{(\AA)} & \colhead{(\AA)} &
\colhead{(mag)} & \colhead{(\AA)} & \colhead{} & \colhead{(dex)} 
}
\startdata
5001  & 1.02 &$-$0.003 &  0.090 & 2.38 & 0.96 & 1.56 & 1.14 &  0.021 & 0.58 & 36.2   & $-$0.95\nl
5002  & 2.32 &   0.035 &  0.085 & 2.04 & 1.89 & 1.74 & 0.40 &  0.004 & 1.89 & 25.7   & $-$1.22\nl
5012  & 2.01 &   0.145 &  0.306 & 5.06 & 3.25 & 3.02 & 5.88 &  0.040 & 1.81 & 23.0   & $+$0.11\nl
5015  & 2.61 &   0.021 &  0.089 & 1.70 & 1.24 & 0.96 & 1.53 &$-$0.013 & 2.72 & 53.9  & $-$1.14\nl
5020  & 1.77 &   0.014 &  0.081 & 1.17 & 2.11 & 1.09 & 1.76 &$-$0.015 & ...~ & 22.1  & $-$1.14\nl
5021  & 0.89 &   0.081 &  0.220 & 4.13 & 2.22 & 1.43 & 3.73 &  0.023 & 1.78 & 24.3   & $-$0.23\nl
5024  & 2.61 &   0.023 &  0.065 & 1.38 & 0.99 & 1.07 & 1.02 &  0.007 & ...~ & 50.8   & $-$1.41\nl
5025  & 1.72 &   0.024 &  0.094 & 2.55 & 1.47 & 1.63 & 1.28 &  0.024 & ...~ & 24.2   & $-$0.83\nl
5026  & 1.67 &   0.033 &  0.103 & 2.86 & 1.41 & 0.55 & 0.75 &  0.013 & 2.13 & 51.0   & $-$0.91\nl
5028  & 1.73 &   0.129 &  0.324 & 5.42 & 3.03 & 2.71 & 6.35 &  0.046 & 1.12 & 56.9   & $+$0.11\nl
58    & 1.54 &$-$0.026 &$-$0.006& 0.53 & 0.43 & 1.21 & 0.60 &$-$0.030 & 1.64 & 46.3  & $-$2.13\nl
66    & 2.06 &$-$0.037 &  0.017 & 1.73 & 1.26 & 1.00 & 1.51 &  0.007 & 3.52 & 33.9   & $-$1.10\nl
101   & 1.28 &$-$0.033 &  0.056 & 1.64 & 0.32 & 0.37 & 1.39 &  0.015 & 2.08 & 26.7   & $-$1.33\nl
107   & 1.41 &$-$0.008 &  0.113 & 3.23 & 2.12 & 1.73 & 2.35 &  0.025 & 2.98 & 44.4   & $-$0.53\nl
141   & 1.94 &   0.004 &  0.067 & 2.23 & 2.44 & 1.58 & 2.51 &  0.021 & 1.72 & 46.8   & $-$0.72\nl
176   & 1.67 &   0.010 &  0.150 & 2.54 & 1.80 & 1.77 & 1.53 &  0.015 & 2.45 & 33.7   & $-$0.80\nl
177   & 1.94 &$-$0.048 &  0.112 & 2.88 & 2.43 & 2.39 & 0.70 &  0.033 & 3.21 & 30.9   & $-$0.80\nl
186   & 2.17 &$-$0.035 &  0.071 & 1.71 & 0.71 & 0.62 & 1.17 &  0.012 & 2.98 & 37.9   & $-$1.29\nl
191   & 1.26 &$-$0.024 &$-$0.008& 1.51 & 1.06 & 0.22 & 0.81 &$-$0.005 & 1.18 & 27.6  & $-$1.48\nl
248   & 2.36 &$-$0.026 &  0.086 & 2.22 & 2.42 & 1.45 & 1.50 &  0.011 & 3.05 & 51.3   & $-$0.87\nl
279   & 1.55 &$-$0.026 &  0.077 & 2.45 & 1.53 & 1.59 & 1.48 &  0.020 & 2.67 & 68.6   & $-$0.83\nl
280   & 2.19 &   0.060 &  0.159 & 3.19 & 2.03 & 0.71 & 1.45 &  0.019 & 0.85 & 31.5   & $-$0.68\nl
290   & 1.70 &   0.066 &  0.201 & 3.68 & 2.21 & 1.78 & 1.06 &  0.038 & 1.56 & 37.9   & $-$0.61\nl
292   & 0.95 &$-$0.009 &  0.042 & 2.13 & 0.69 &$-$0.13 & 1.59 &$-$0.003 & ...~ &26.9 & $-$1.10\nl
307   & 2.76 &$-$0.038 &  0.059 & 1.85 & 2.02 & 1.37 & 1.49 &  0.008 & 2.89 & 41.4   & $-$0.99\nl
311   & 1.38 &$-$0.023 &  0.108 & 3.03 & 1.31 & 2.17 & 1.26 &  0.034 & 2.53 & 28.1   & $-$0.72\nl
313   & 0.33 &$-$0.035 &  0.062 & 2.84 & 1.77 & 0.67 & 0.55 &  0.014 & 1.69 & 22.3   & $-$0.91\nl
314   & 1.88 &   0.016 &  0.146 & 4.15 & 2.69 & 2.92 & 2.69 &  0.047 & 1.96 & 65.8   & $-$0.34\nl
321   & 2.36 &$-$0.073 &$-$0.028& 0.96 & 1.38 & 0.26 & 0.73 &$-$0.003 & 3.24 & 62.3  & $-$1.75\nl
323   & 2.62 &$-$0.060 &$-$0.049& 0.31 & 1.60 & 0.93 & 0.52 &$-$0.019 & 2.67 & 43.2  & $-$2.13\nl
324   & 1.09 &$-$0.019 &  0.091 & 2.76 & 1.90 & 1.28 & 1.62 &  0.002 & 2.84 & 34.5   & $-$0.76\nl
330   & 2.41 &   0.009 &  0.123 & 3.06 & 2.68 & 1.95 & 1.85 &  0.027 & 2.45 & 33.6   & $-$0.61\nl
348   & 2.10 &$-$0.030 &  0.009 & 1.43 & 2.00 & 1.35 & 1.07 &  0.010 & 1.86 & 83.6   & $-$1.25\nl
350   & 2.79 &$-$0.035 &  0.026 & 2.30 & 1.83 & 1.58 & 0.75 &  0.005 & 3.07 & 36.6   & $-$1.02\nl
376   & 2.84 &$-$0.062 &$-$0.027& 1.18 & 0.86 & 1.13 & 0.83 &  0.010 & 3.33 & 52.6   & $-$1.52\nl
417   & 1.60 &   0.003 &  0.140 & 3.04 & 2.12 & 1.76 & 1.97 &  0.028 & 2.56 &109.9   & $-$0.61\nl
418   & 2.02 &$-$0.051 &$-$0.026& 1.26 & 0.79 & 0.85 & 1.30 &  0.014 & 2.67 & 38.3   & $-$1.41\nl
421   & 1.86 &   0.020 &  0.184 & 3.13 & 2.92 & 2.52 & 3.12 &  0.026 & ...~ & 50.1   & $-$0.42\nl
423   & 2.58 &   0.050 &  0.176 & 3.98 & 2.12 & 2.07 & 2.94 &  0.025 & 1.87 & 45.0   & $-$0.34\nl
442   & 1.93 &$-$0.038 &  0.007 & 1.67 & 1.82 & 1.26 & 1.10 &  0.014 & 2.83 & 61.5   & $-$1.18\nl
453   & 2.97 &$-$0.038 &  0.070 & 1.60 & 2.54 & 1.52 & 1.80 &  0.008 & 3.39 & 33.3   & $-$0.99\nl
490   & 1.41 &   0.094 &  0.262 & 5.29 & 3.22 & 3.01 & 6.77 &  0.043 & 2.10 & 70.3   & $+$0.11\nl
491   & 1.83 &$-$0.036 &  0.019 & 1.03 & 1.84 & 0.58 & 0.83 &  0.019 & 1.99 & 47.2   & $-$1.56\nl
492   & 2.69 &$-$0.000 &  0.131 & 2.64 & 2.23 & 1.80 & 2.60 &  0.006 & 2.46 & 24.3   & $-$0.61\nl
519   & 1.49 &   0.012 &  0.127 & 3.06 & 3.04 & 0.26 & 2.15 &  0.023 & 1.68 & 29.2   & $-$0.61\nl
526   & 0.73 &   0.038 &  0.194 & 4.45 & 3.07 & 0.59 & 3.53 &  0.021 & ...~ & 44.0   & $-$0.23\nl
537   & 2.98 &$-$0.048 &  0.033 & 2.06 & 1.98 &$-$0.48 & 0.49 &  0.036 & 3.08 & 23.0 & $-$1.37\nl
571   & 1.56 &$-$0.011 &  0.102 & 2.70 & 2.14 & 2.14 & 1.38 &  0.007 & 3.04 & 29.8   & $-$0.76\nl
579   & 2.37 &   0.011 &  0.034 & 0.81 & 0.57 & 0.74 & 0.82 &  0.009 & ...~ & 40.9   & $-$1.90\nl
581   & 1.88 &   0.020 &  0.094 & 1.79 & 1.67 & 2.38 & 1.69 &  0.024 & 2.23 & 43.0   & $-$0.95\nl
588   & 2.22 &$-$0.057 &  0.043 & 1.92 & 1.44 & 1.01 & 1.24 &  0.003 & 3.12 & 53.1   & $-$1.10\nl
602   & 1.07 &$-$0.030 &  0.077 & 2.86 & 2.13 & 1.59 & 1.62 &  0.018 & 2.79 & 46.8   & $-$0.72\nl
611   & 2.74 &$-$0.014 &  0.087 & 1.01 & 1.37 & 1.41 & 0.44 &$-$0.001 & ...~ &37.0   & $-$1.59\nl
614   & 3.89 &$-$0.026 &  0.061 & 1.85 & 1.96 & 0.19 & 1.13 &  0.011 & 2.72 & 25.1   & $-$1.18\nl
647   & 2.10 &   0.025 &  0.137 & 2.10 & 2.39 & 1.59 & 1.55 &  0.009 & ...~ & 51.9   & $-$0.87\nl
649   & 2.04 &$-$0.043 &  0.005 & 1.55 & 1.47 & 1.31 & 0.99 &  0.005 & 3.09 & 36.2   & $-$1.29\nl
651   & 1.50 &   0.123 &  0.304 & 5.19 & 2.29 & 2.54 & 4.83 &  0.037 & ...~ & 25.9   & $+$0.04\nl
664   & 1.50 &   0.033 &  0.112 & 2.15 & 2.26 & 1.56 & 2.23 &  0.032 & 2.09 & 24.6   & $-$0.76\nl
672   & 2.19 &$-$0.014 &  0.038 & 1.26 & 0.89 & 0.88 & 0.85 &$-$0.013 & 2.97 &52.4   & $-$1.52\nl
679   & 0.59 &   0.011 &  0.150 & 3.66 & 2.54 & 1.53 & 3.00 &  0.034 & 2.49 & 36.1   & $-$0.38\nl
680   & 2.43 &$-$0.044 &  0.037 & 0.74 & 1.52 & 1.56 & 1.18 &  0.003 & 2.57 & 40.9   & $-$1.48\nl
682   & 2.22 &$-$0.047 &  0.027 & 1.15 & 1.29 &$-$0.05 & 0.30 &  0.005 & 2.67 &30.2  & $-$1.75\nl
697   & 3.21 &   0.046 &  0.159 & 4.88 & 4.01 & 3.26 & 2.40 &  0.035 & 2.32 & 23.1   & $-$0.27\nl
715   & 1.47 &   0.018 &  0.154 & 2.84 & 1.67 & 1.18 & 1.39 &  0.014 & 2.26 & 27.4   & $-$0.76\nl
723   & 1.84 &$-$0.025 &  0.042 & 0.90 & 0.09 & 0.04 & 0.33 &$-$0.000 & 1.88 & 31.0  & $-$2.24\nl
741   & 1.38 &   0.088 &  0.209 & 3.77 & 2.20 & 2.34 & 3.33 &  0.043 & 1.53 & 42.7   & $-$0.30\nl
746   & 1.77 &   0.003 &  0.155 & 2.94 & 1.98 & 1.96 & 2.44 &  0.024 & 2.63 & 78.4   & $-$0.57\nl
750   & 0.93 &   0.021 &  0.069 & 1.85 & 1.12 & 0.91 & 0.97 &  0.002 & ...~ & 40.9   & $-$1.25\nl
770   & 2.19 &   0.009 &  0.136 & 2.19 & 1.51 & 1.47 & 2.37 &  0.011 & ...~ & 45.5   & $-$0.76\nl
784   & 1.58 &$-$0.025 &  0.053 & 2.57 & 2.07 & 2.00 & 2.13 &  0.001 & 1.77 & 64.6   & $-$0.68\nl
787   & 2.54 &$-$0.014 &  0.035 & 1.69 & 2.10 & 1.19 & 1.01 &  0.008 & 2.65 & 24.9   & $-$1.18\nl
796   & 2.80 &$-$0.065 &$-$0.001& 1.22 & 1.16 & 0.57 & 0.76 &$-$0.004 & 1.41 & 41.4  & $-$1.56\nl
798   & ...~ &$-$0.036 &  0.015 & 1.38 & 1.11 & 0.29 & 1.12 &$-$0.002 & 0.97 & 36.3  & $-$1.44\nl
809   & 2.07 &$-$0.033 &  0.013 & 1.34 & 1.46 & 1.67 & 0.92 &  0.015 & 1.70 & 44.5   & $-$1.33\nl
811   & 2.19 &   0.000 &  0.009 & 0.18 &$-$0.20 & 0.21 &$-$0.27 &$-$0.001& 2.92 &40.3& $-$2.24\nl
824   & 2.11 &   0.034 &  0.110 & 2.07 & 3.94 & 1.53 & 1.76 &$-$0.002 & 1.95 & 32.5  & $-$0.80\nl
838   & 2.75 &$-$0.025 &  0.017 & 2.18 & 1.24 & 0.83 & 1.65 &  0.003 & 1.98 & 33.4   & $-$0.95\nl
849   & 2.32 &$-$0.006 &  0.105 & 3.73 & 2.63 & 1.83 & 2.47 &  0.038 & 2.55 & 32.8   & $-$0.46\nl
868   & 1.71 &$-$0.027 &  0.085 & 2.59 & 3.23 & 1.63 & 0.73 &  0.023 & 3.07 & 25.5   & $-$0.83\nl
871   & 1.31 &   0.015 &  0.113 & 3.19 & 2.89 & 2.17 & 2.75 &  0.011 & 2.41 & 27.3   & $-$0.49\nl
887   & 1.29 &$-$0.021 &  0.084 & 1.72 & 2.00 & 1.62 & 1.68 &  0.020 & 2.75 & 33.7   & $-$0.99\nl
892   & 1.62 &$-$0.016 &  0.062 & 3.24 & 1.75 & 2.97 & 2.24 &  0.032 & 2.18 & 24.6   & $-$0.53\nl
902   & 0.57 &   0.151 &  0.361 & 5.06 & 3.98 & 3.01 & 7.63 &  0.048 & 2.06 & 24.1   & $+$0.11\nl
910   & 2.72 &$-$0.043 &$-$0.042& 0.25 & 0.47 & 1.00 & 0.66 &$-$0.009 & 2.58 & 39.1  & $-$2.24\nl
917   & 2.37 &$-$0.028 &  0.091 & 2.36 & 1.42 & 1.55 & 1.66 &  0.023 & 2.62 & 22.7   & $-$0.83\nl
928   & 1.96 &$-$0.032 &  0.050 & 1.32 & 1.26 & 1.08 & 1.07 &  0.005 & 2.79 & 65.0   & $-$1.37\nl
937   & 2.12 &$-$0.028 &  0.005 & 1.28 & 1.11 & 1.13 & 0.84 &$-$0.008 & 3.23 & 28.9  & $-$1.44\nl
941   & 2.11 &$-$0.037 &  0.020 & 0.85 & 1.26 & 0.41 & 1.54 &$-$0.000 & 2.54 & 46.7  & $-$1.59\nl
946   & 1.66 &   0.077 &  0.273 & 3.74 & 4.48 & 3.25 & 2.83 &  0.019 & ...~ & 29.6   & $-$0.30\nl
952   & 1.93 &$-$0.049 &  0.032 & 1.38 & 1.42 & 0.94 & 1.21 &  0.005 & 3.17 & 32.2   & $-$1.29\nl
965   & 2.03 &   0.066 &  0.156 & 2.87 & 1.80 & 1.44 & 2.09 &  0.029 & ...~ & 52.5   & $-$0.65\nl
968   & 2.38 &$-$0.006 &  0.064 & 1.95 & 1.02 & 1.95 & 1.01 &  0.013 & 2.87 & 33.5   & $-$1.14\nl
970   & 2.06 &$-$0.040 &  0.094 & 2.50 & 2.19 & 2.42 & 1.86 &  0.035 & 2.99 & 32.5   & $-$0.72\nl
991   & 1.68 &$-$0.029 &  0.053 & 1.52 & 1.41 & 1.31 & 0.92 &$-$0.003 & 2.67 & 44.5  & $-$1.29\nl
992   & 2.82 &$-$0.044 &  0.057 & 2.38 & 1.01 & 2.46 & 0.90 &  0.013 & 3.96 & 25.5   & $-$0.95\nl
1007  & 1.25 &   0.032 &  0.189 & 3.20 & 1.63 & 2.57 & 2.01 &  0.028 & ...~ & 58.2   & $-$0.57\nl
1010  & 0.98 &   0.049 &  0.241 & 4.37 & 2.45 & 2.15 & 3.61 &  0.043 & ...~ & 38.3   & $-$0.19\nl
1016  & 1.14 &   0.159 &  0.398 & 5.65 & 3.38 & 3.42 & 7.46 &  0.037 & ...~ & 37.1   & $+$0.11\nl
1019  & 2.02 &$-$0.039 &  0.056 & 1.89 & 0.87 & 1.50 & 1.12 &  0.001 & 3.09 & 74.9   & $-$1.18\nl
1032  & 3.29 &$-$0.007 &  0.127 & 2.47 & 2.09 & 1.33 & 1.42 &  0.012 & 2.57 & 41.3   & $-$0.83\nl
1034  & 2.09 &   0.057 &  0.135 & 2.26 & 1.38 & 1.63 & 1.75 &  0.017 & ...~ & 39.3   & $-$0.83\nl
1044  & 2.24 &$-$0.070 &  0.015 & 2.00 & 1.96 & 0.89 & 1.39 &  0.002 & 2.94 & 44.4   & $-$1.02\nl
1049  & 1.68 &   0.015 &  0.169 & 3.87 & 3.00 & 2.20 & 2.62 &  0.019 & ...~ & 24.6   & $-$0.38\nl
1060  & 1.64 &   0.005 &  0.101 & 1.08 & 1.18 & 2.02 & 1.26 &$-$0.013 & 1.73 & 22.7  & $-$1.29\nl
1091  & 1.73 &$-$0.011 &  0.100 & 1.74 & 1.77 & 0.39 & 2.15 &$-$0.009 & 2.72 & 29.7  & $-$0.99\nl
1093  & ...~ &$-$0.004 &  0.073 & 2.21 & 2.06 & 1.44 & 1.46 &  0.016 & 0.64 & 48.5   & $-$0.87\nl
1101  & 1.75 &$-$0.013 &  0.127 & 1.47 & 2.60 & 1.97 & 1.99 &  0.019 & 2.28 & 28.6   & $-$0.95\nl
1110  & 2.90 &$-$0.062 &  0.011 & 1.45 & 1.21 & 0.88 & 1.27 &  0.001 & 3.01 & 22.1   & $-$1.29\nl
1113  & 2.16 &$-$0.011 &  0.044 & 0.39 &$-$0.08 & 0.64 & 1.11 &  0.019 & ...~ &25.9  & $-$2.17\nl
1116  & 2.51 &   0.015 &  0.147 & 2.44 & 2.56 & 0.58 & 1.67 &  0.006 & ...~ & 28.5   & $-$0.80\nl
1119  & 2.12 &$-$0.044 &  0.109 & 1.97 & 1.40 & 1.92 & 2.06 &$-$0.006 & ...~ & 23.5  & $-$0.87\nl
1120  & 2.29 &$-$0.009 &  0.090 & 1.45 & 1.72 & 1.46 & 0.80 &$-$0.007 & ...~ & 26.0  & $-$1.33\nl
1155  & 1.28 &   0.060 &  0.240 & 4.02 & 2.55 & 2.23 & 3.26 &  0.024 & ...~ & 43.0   & $-$0.30\nl
1157  & 1.45 &$-$0.033 &  0.022 & 1.52 & 1.84 & 1.77 & 1.22 &  0.011 & 3.18 & 46.0   & $-$1.14\nl
1158  & 1.87 &$-$0.008 &  0.118 & 1.31 & 0.98 & 1.01 & 1.35 &  0.005 & ...~ & 33.4   & $-$1.33\nl
1167  & 2.23 &$-$0.041 &  0.027 & 0.99 & 0.56 & 0.57 & 0.46 &  0.001 & ...~ & 23.9   & $-$1.86\nl
1180  & 1.98 &   0.013 &  0.117 & 1.73 & 1.62 & 2.11 & 1.49 &  0.020 & ...~ & 65.8   & $-$1.02\nl
1181  & 0.67 &   0.020 &  0.169 & 2.99 & 1.78 & 1.73 & 1.70 &  0.012 & ...~ & 27.3   & $-$0.68\nl
1205  & 2.13 &$-$0.060 &  0.007 & 0.90 & 0.58 & 0.79 & 1.94 &  0.006 & 2.60 & 26.1   & $-$1.37\nl
1238  & 2.36 &   0.007 &  0.085 & 2.38 & 0.98 & 1.85 & 1.10 &  0.022 & 2.60 & 28.5   & $-$0.95\nl
1240  & 1.08 &   0.004 &  0.152 & 4.39 & 3.08 & 2.69 & 2.24 &  0.024 & 2.52 & 37.2   & $-$0.38\nl
1247  & 2.39 &$-$0.023 &  0.072 & 2.54 & 1.60 & 1.01 & 2.54 &  0.013 & 3.07 & 37.7   & $-$0.68\nl
1290  & 2.43 &$-$0.016 &  0.142 & 2.60 & 2.14 & 1.04 & 1.25 &  0.007 & 2.87 & 24.3   & $-$0.83\nl
1293  & 1.67 &$-$0.011 &  0.051 & 2.18 & 2.27 & 1.19 & 2.11 &  0.032 & 1.59 & 32.3   & $-$0.80\nl
1309  & 1.69 &$-$0.007 &  0.089 & 2.79 & 2.49 & 1.83 & 2.65 &  0.018 & 2.61 & 50.1   & $-$0.57\nl
1336  & 2.46 &$-$0.044 &  0.091 & 1.89 & 1.65 & 1.27 & 1.50 &  0.008 & 2.80 & 50.8   & $-$1.02\nl
1344  & 2.34 &$-$0.018 &  0.080 & 2.84 & 1.92 & 1.23 & 2.05 &  0.018 & 2.63 & 65.3   & $-$0.68\nl
1351  & 1.70 &$-$0.009 &  0.089 & 3.20 & 2.12 & 1.85 & 2.00 &  0.004 & 2.06 & 53.9   & $-$0.61\nl
1353  & 2.08 &$-$0.033 &  0.037 & 0.60 & 1.94 & 1.65 & 0.87 &  0.002 & 3.45 & 25.5   & $-$1.59\nl
1370  & 1.91 &$-$0.010 &  0.131 & 3.25 & 2.51 & 1.97 & 2.62 &  0.022 & 2.50 & 83.6   & $-$0.49\nl
1382  & 2.00 &$-$0.062 &$-$0.038& 0.98 & 0.73 & 0.42 & 0.53 &$-$0.002 & 2.42 & 38.0  & $-$1.86\nl
1431  & 2.54 &$-$0.086 &$-$0.070& 1.01 &$-$1.53 &$-$0.68 & 0.86 &$-$0.009 & 2.90&29.4& $-$2.24\nl
1449  & 2.42 &$-$0.056 &$-$0.017& 1.67 & 1.26 & 0.15 & 1.03 &  0.014 & 1.52 & 31.8   & $-$1.37\nl
1463  & 1.42 &$-$0.075 &  0.002 & 0.87 & 1.59 & 1.43 & 1.71 &  0.015 & 2.45 & 36.3   & $-$1.25\nl
1479  & 1.96 &$-$0.029 &  0.071 & 1.59 & 1.20 & 1.73 & 0.82 &  0.011 & 2.93 & 41.8   & $-$1.29\nl
1481  & 2.03 &$-$0.031 &  0.048 & 1.58 & 1.59 & 1.28 & 2.00 &  0.007 & 2.83 & 38.3   & $-$1.02\nl
1483  & 2.65 &$-$0.096 &$-$0.016& 0.54 & 0.64 & 2.21 & 1.39 &  0.010 & 2.11 & 23.2   & $-$1.63\nl
1504  & 2.09 &$-$0.068 &$-$0.061& 0.52 & 0.29 & 0.51 & 0.54 &  0.018 & 1.88 & 52.7   & $-$2.24\nl
1514  & 1.86 &   0.036 &  0.166 & 4.54 & 2.55 & 2.61 & 2.83 &  0.028 & 1.30 & 48.2   & $-$0.27\nl
1538  & 1.63 &   0.004 &  0.094 & 2.86 & 2.46 & 2.19 & 2.08 &  0.032 & 1.83 & 70.1   & $-$0.61\nl
1540  & 2.05 &   0.023 &  0.170 & 3.63 & 2.86 & 2.24 & 1.94 &  0.014 & 2.31 & 28.1   & $-$0.49\nl
1548  & 2.39 &   0.011 &  0.144 & 2.36 & 2.53 & 2.10 & 1.20 &  0.015 & 2.02 & 61.6   & $-$0.83\nl
1563  & 2.60 &$-$0.033 &  0.057 & 2.28 & 2.35 & 0.72 & 0.97 &$-$0.005 & 2.91 & 25.6  & $-$0.99\nl
1565  & 2.60 &$-$0.057 &  0.030 & 1.26 & 1.58 & 1.26 & 0.70 &  0.003 & 3.15 & 34.4   & $-$1.44\nl
1594  & 2.09 &$-$0.021 &  0.058 & 1.73 & 1.68 & 1.33 & 1.34 &  0.021 & 2.42 & 37.1   & $-$1.10\nl
1615  & 2.13 &$-$0.020 &  0.049 & 2.26 & 1.61 & 1.67 & 1.22 &  0.012 & 2.67 & 55.8   & $-$0.95\nl
1617  & 1.61 &$-$0.032 &  0.054 & 2.18 & 1.97 & 1.46 & 1.21 &  0.004 & 2.99 & 74.6   & $-$0.95\nl
1631  & 1.54 &$-$0.029 &  0.049 & 1.14 & 2.61 & 1.26 & 1.30 &  0.004 & 3.23 & 38.4   & $-$1.22\nl
1664  & 1.58 &$-$0.030 &  0.085 & 2.58 & 1.96 & 0.66 & 1.81 &  0.026 & 2.82 & 51.6   & $-$0.76\nl
1709  & 1.22 &   0.015 &  0.177 & 2.75 & 2.14 & 2.46 & 1.68 &  0.014 & 2.73 & 22.9   & $-$0.68\nl
\enddata
\end{deluxetable}

\begin{deluxetable}{lrrrrrrrrrc}
\scriptsize
\tablewidth{0pt}
\tablecaption{Index Measurements for Galactic Globular Clusters}
\tablehead{
\colhead{GC} & \colhead{H$\beta$} & \colhead{Mg$_1$} & \colhead{Mg$_2$} & 
\colhead{Mg $b$} & \colhead{Fe5270} & \colhead{Fe5335} & \colhead{NaD} &
\colhead{TiO$_1$} & \colhead{H$\alpha$} & \colhead{[Fe/H]$_Z$~~$\pm$~}
\\
\colhead{} & \colhead{(\AA)} & \colhead{(mag)} & \colhead{(mag)} & 
\colhead{(\AA)} & \colhead{(\AA)} & \colhead{(\AA)} & \colhead{(\AA)} &
\colhead{(mag)} & \colhead{(\AA)} & \colhead{(dex)\phantom{~~$\pm$~}} 
}
\startdata
M92   & 2.53 & 0.002 & 0.028 & 0.57 & 0.53 & 0.37 & 0.67 & 0.001 & 2.35 & $-$2.24~~~0.08 \nl
M13   & 2.40 & 0.007 & 0.048 & 0.95 & 0.92 & 0.71 & 0.90 & 0.002 & 2.29 & $-$1.65~~~0.06 \nl
M4    & 2.82 & 0.030 & 0.095 & 1.59 & 1.28 & 0.97 & 1.67 & 0.009 & 2.54 & $-$1.28~~~0.10 \nl
N6171 & 2.00 & 0.041 & 0.119 & 1.69 & 1.60 & 1.47 & 1.80 & 0.016 & 2.90 & $-$0.99~~~0.06 \nl
N6539 & 1.39 & 0.085 & 0.195 & 2.86 & 2.30 & 1.77 & 3.76 & 0.043 & 1.52 & $-$0.66~~~0.15 \nl
N6356 & 1.62 & 0.070 & 0.169 & 3.09 & 2.00 & 1.69 & 3.00 & 0.029 & ...~ & $-$0.62~~~0.20 \nl
M71   & 1.29 & 0.071 & 0.158 & 2.84 & 1.80 & 1.64 & 2.25 & 0.022 & ...~ & $-$0.58~~~0.08 \nl
N6760 & 1.34 & 0.096 & 0.228 & 3.44 & 2.52 & 2.01 & 3.08 & 0.050 & 1.26 & $-$0.52~~~0.15 \nl
N6624 & 1.69 & 0.048 & 0.163 & 2.94 & 2.09 & 1.78 & 2.20 & 0.035 & 1.63 & $-$0.35~~~0.15 \nl
N6553 & 1.63 & 0.110 & 0.249 & 3.88 & 3.11 & 2.51 & 3.40 & 0.044 & 1.26 & $-$0.29~~~0.11 \nl
N6440 & 1.34 & 0.103 & 0.227 & 3.31 & 2.50 & 1.96 & 3.94 & 0.034 & 1.48 & $-$0.26~~~0.15 \nl
N6528 & 1.80 & 0.097 & 0.247 & 3.89 & 2.96 & 2.45 & 4.93 & 0.046 & 1.29 & $+$0.12~~~0.21 \nl
\enddata
\end{deluxetable}

\begin{deluxetable}{ccccc}
\scriptsize
\tablewidth{0pt}
\tablecaption{Age Estimates by Metallicity for the M87 Globular Clusters}
\tablehead{
\colhead{} & \colhead{Median} & \colhead{Mean} & \colhead{Median} & \colhead{} \\
\colhead{Bin} & \colhead{[Fe/H]$_Z$} & \colhead{Age} 
& \colhead{Age} & \colhead{$\sigma${$_{\rm Age}$}}
\\
\colhead{} & \colhead{(dex)} & \colhead{(Gyr)} & \colhead{(Gyr)} & \colhead{(Gyr)}
}
\startdata
 1  & $-$1.75 & 14.6 & 14.0 & 1.7 \\
 2  & $-$1.29 & 13.8 & 13.0 & 1.7 \\
 3  & $-$1.02 & 12.6 & 13.0 & 2.0 \\
 4  & $-$0.84 & 11.8 & 12.0 & 1.9 \\
 5  & $-$0.68 & 12.6 & 13.0 & 1.5 \\
 6  & $-$0.30 & 13.6 & 13.0 & 3.1 \\
\enddata
\end{deluxetable}

%
\begin{deluxetable}{lll|lll}
\tablewidth{0pt}
\scriptsize
\tablecaption{Comparison of Mean [Fe/H] for M87 Globular Clusters}
\tablehead{
   \colhead{Reference}
 & \colhead{$<$[Fe/H]$>$}
 & \colhead{$\sigma$}
 & \colhead{Reference}
 & \colhead{$<$[Fe/H]$>$}
 & \colhead{$\sigma$}
\\
   \colhead{}
 & \colhead{(dex)}
 & \colhead{(dex)}
 &  \colhead{}
 & \colhead{(dex)}
 & \colhead{(dex)}
}
\startdata
Photometry & (ground based) & & Spectroscopy \\
Cohen (1988) & $-1.0$ & ... & Mould et al (1990) & $-1.0$ & ... \\
Couture et al (1990) & $-1.1$ & 0.6 & Brodie \& Huchra (1991) & $-0.9$ & 
... \\
Lee \& Geisler (1993) & $-0.85$ & 0.65 & This paper & $-0.95$ & 0.5 \\
\enddata
\end{deluxetable}

%
%
\begin{deluxetable}{lrr|lrr}\tablenum{A1}
\tablewidth{0pt}
\scriptsize
\tablecaption{New and Corrected Radial Velocities for M87 Globular Clusters}
\tablehead{
   \colhead{ID Number}
 & \colhead{$v_r$ (Paper~I)}
 & \colhead{$v_r$ (new)}
 & \colhead{ID Number}
 & \colhead{$v_r$ (Paper~I)}
 & \colhead{$v_r$ (new)}
\\
   \colhead{}
 & \colhead{(\kms)}
 & \colhead{(\kms)}
 &  \colhead{}
 & \colhead{(\kms)}
 & \colhead{(\kms)}
}
\startdata
 5001 &  929 & 680  &  5020 & 1646 & 1450 \\ 
 682 & ... & 1300 &   695 & 1869 & 1650 \\
 1119 & 1997 & 1510 &  1353 & 2161 & 1980 \\
\enddata
\end{deluxetable}

\begin{deluxetable}{lcc|lcc|lcc}\tablenum{B1}
\scriptsize
\tablewidth{0pt}
\tablecaption{New Photometry for Globular Clusters in M87}
\tablehead{
\colhead{ID} & \colhead{$U$} & \colhead{$R$} &
\colhead{ID} & \colhead{$U$} & \colhead{$R$} &
\colhead{ID} & \colhead{$U$} & \colhead{$R$} 
}
\startdata
 5001 &  ... &  19.14 &  395  &  ...  & 20.83 &  902 &  22.34 & 20.20 \\
 5002 & 21.29&   ...  &	 423  & 22.84 & 20.81 &  917 &   ...  & 20.73 \\
 5012 &  ... &  20.19 &	 453  &  ...  & 20.53 &  928 &  20.20 & 19.13 \\
 5015 & 20.83&  18.71 &	 519  &  ...  & 20.79 &  968 &   ...  & 20.87 \\
 5020 & 21.45&  19.74 &	 581  &  ...  & 20.27 &  1032&  21.40 & 20.21 \\
 5021 &  ... &  19.95 &	 682  &  ...  & 20.50 &  1067&  22.06 & 19.90 \\
 5024 & 21.22&  19.56 &	 697  & 21.86 & 19.68 &  1091&  22.07 & 20.26 \\
 5025 & 21.61&  19.37 &	 723  &  ...  & 20.69 &  1101&  22.16 & 20.57 \\
 5026 & 20.88&  19.25 &	 746  & 20.99 & 19.17 &  1110&   ...  & 20.73 \\
 5028 & 21.90&  19.19 &	 750  & 21.30 & 19.85 &  1113&   ...  & 20.43 \\
 248  &  ...  & 19.77 &  770  & 21.87 & 20.31 &  1119&    ... & 20.90 \\
 311  & 22.37 & 20.60 &  798  &  ...  & 20.17 &  1180&   20.93& 19.39 \\
 324  &  ...  & 20.34 &  871  & 22.56 & 20.64 &  1483&    ... & 20.87 \\
 330  &  ...  & 20.66 &  892  & 22.49 & 20.25 &   651&    ... & 20.60 \\
\enddata
\end{deluxetable}

\clearpage

\figcaption[ 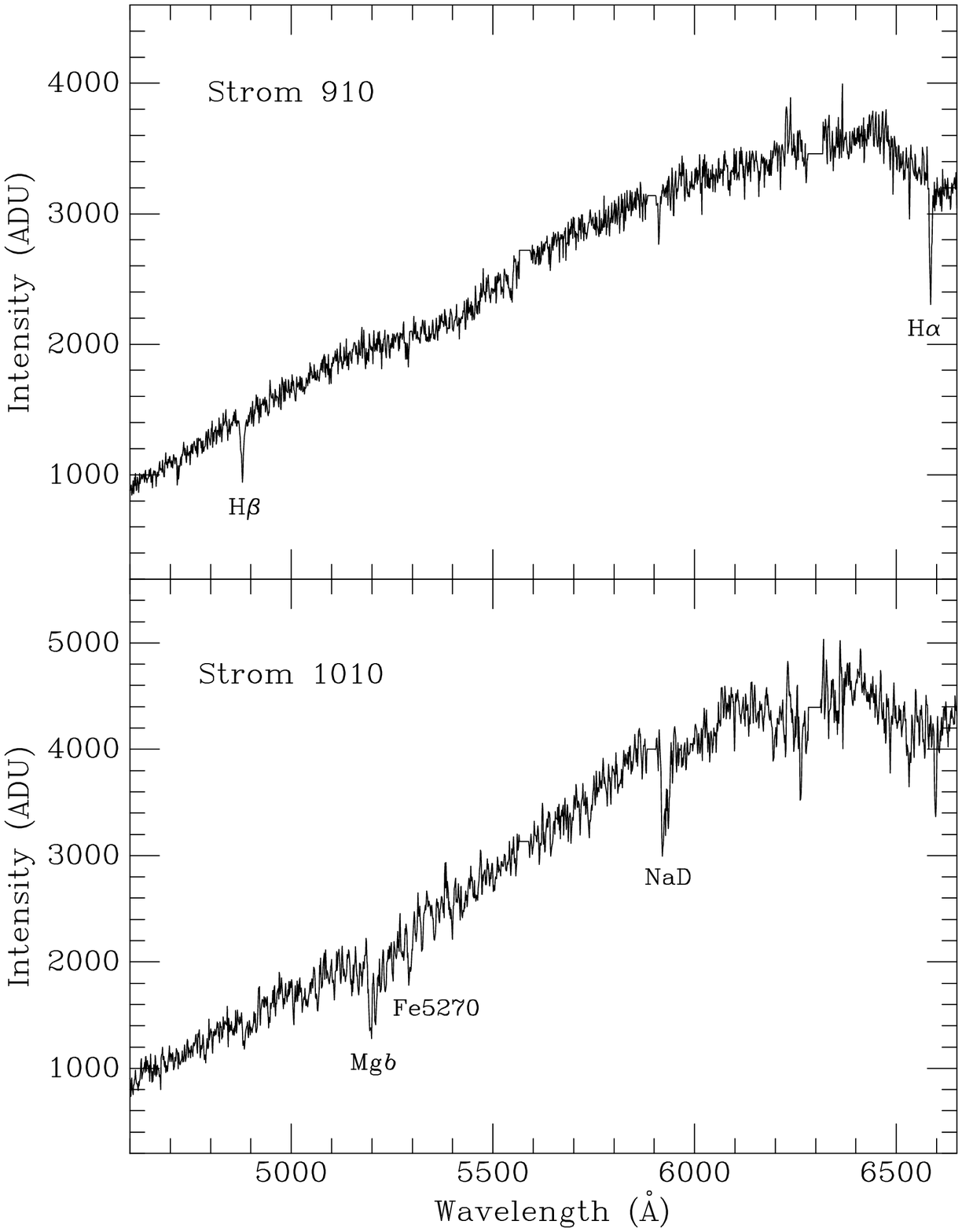]
{A comparison of the spectrum of a metal poor GC in M87 (Strom 910)
with that of a metal rich one (Strom 1010).  Both of these
spectra have $QSNR \approx 40$.}
\label{fig1}

\figcaption[ 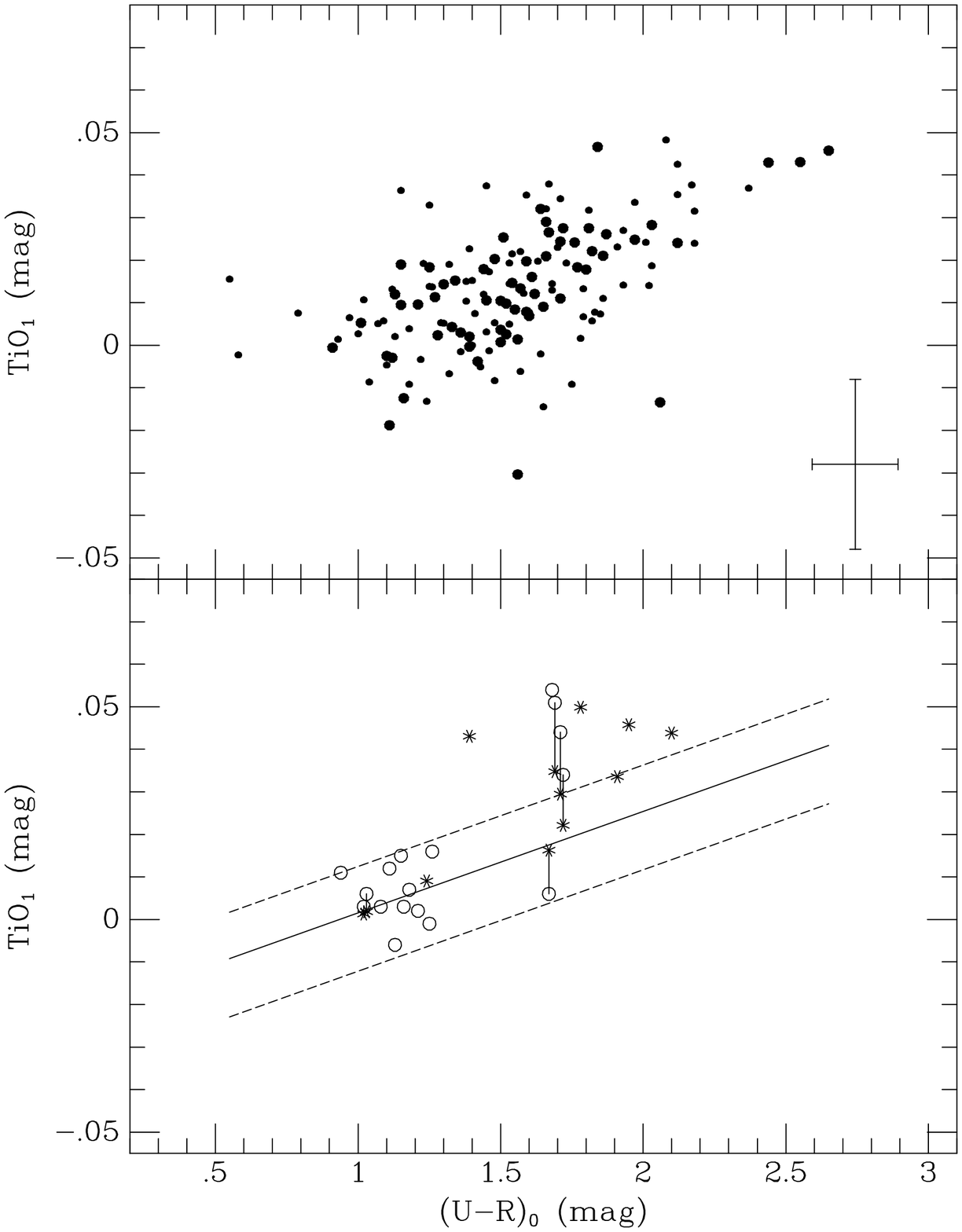]
{The TiO$_1$ index is shown as a function of $(U-R)$ color.
The upper panel shows the data for the M87 GCs, with
objects with $QSNR \ge 40$ indicated by the larger circles.  In the
lower panel the data for the galactic GCs (``*'' our data,
open circles from Burstein et al 1984) is compared with the
``median line'' of the M87 sample.}
\label{fig2}

\figcaption[ 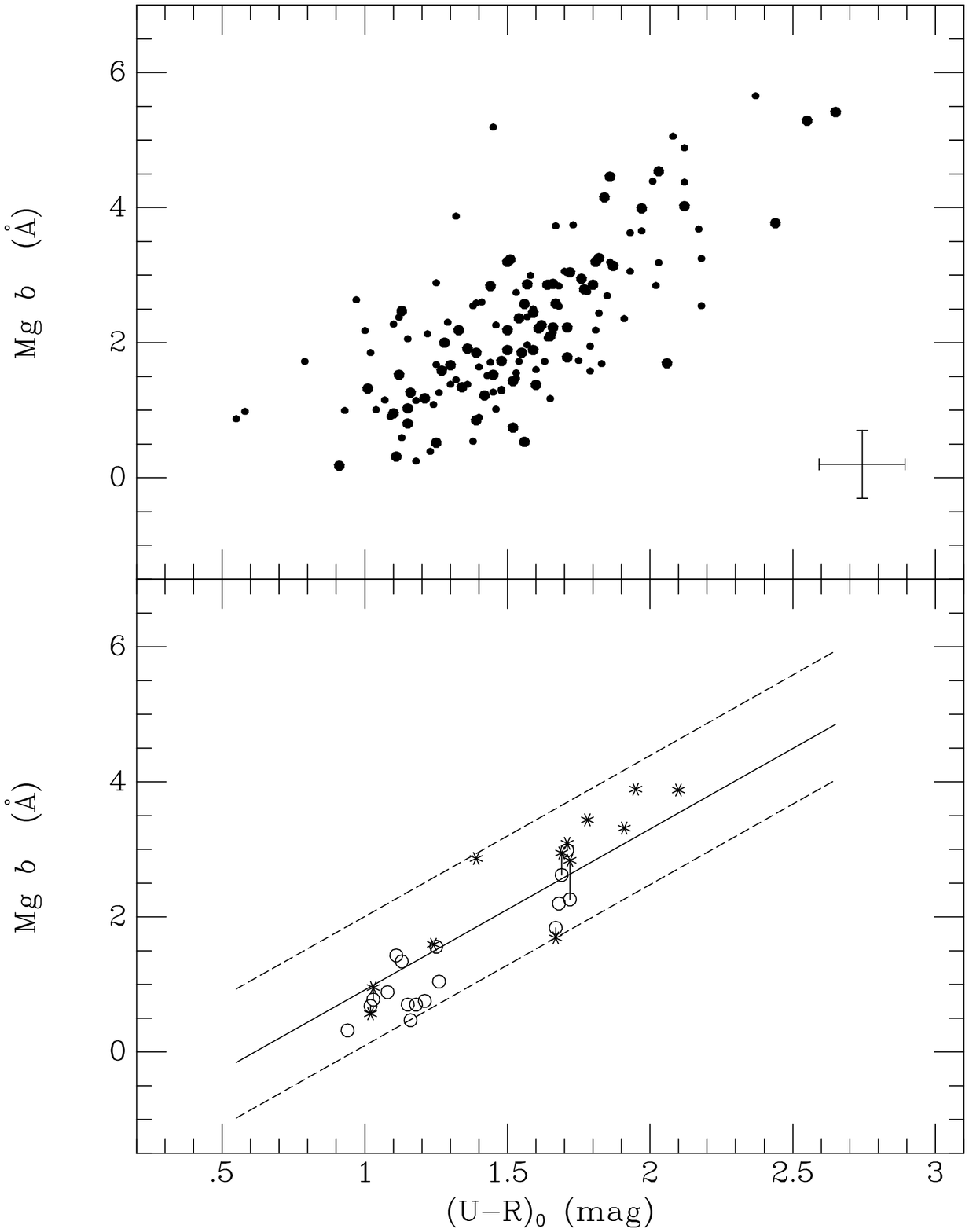]
{The same as Figure 2 for the Mg $b$ index.}
\label{fig3}

\figcaption[ 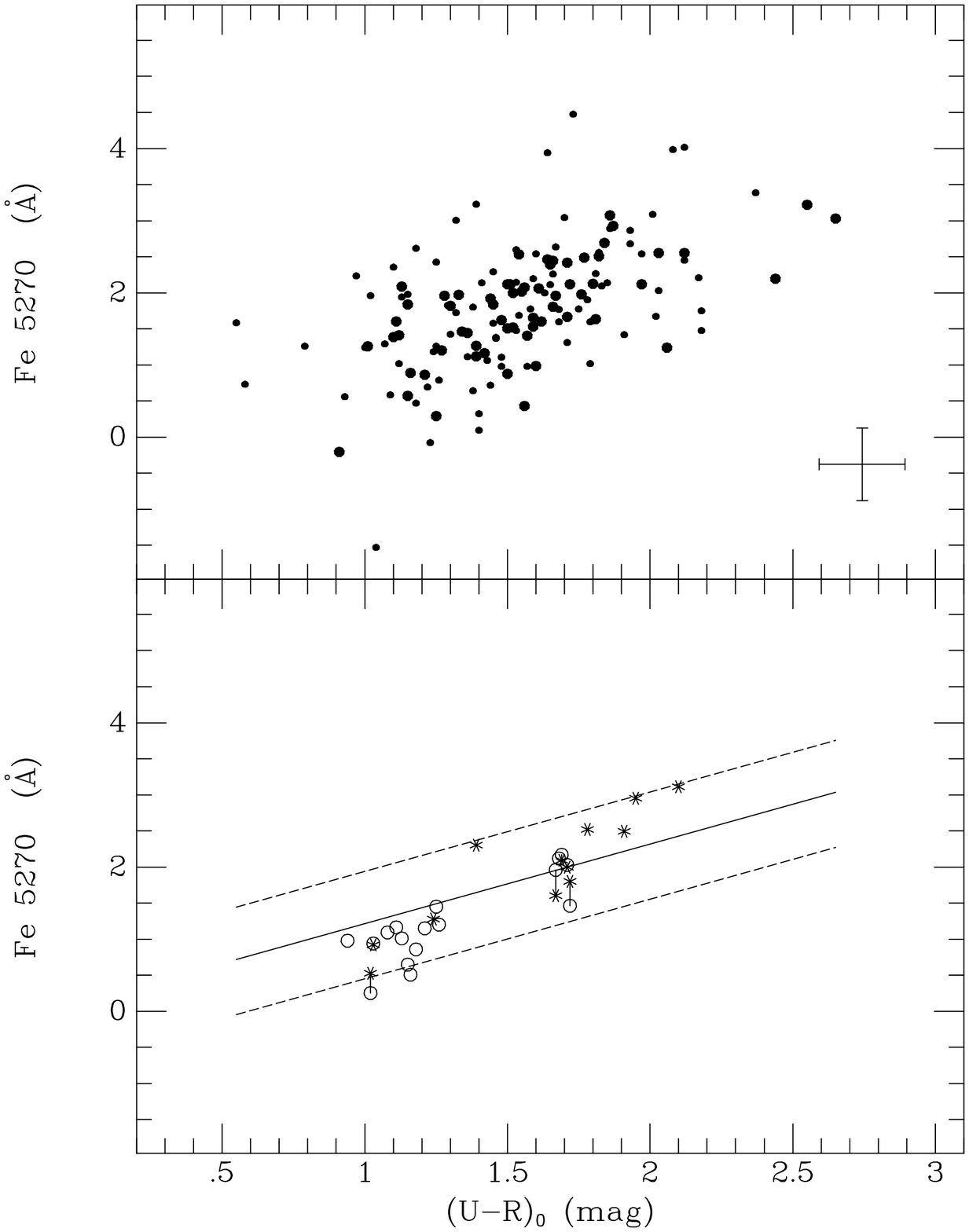]
{The same as Figure 2 for the Fe 5270 index.}
\label{fig4}

\figcaption[ 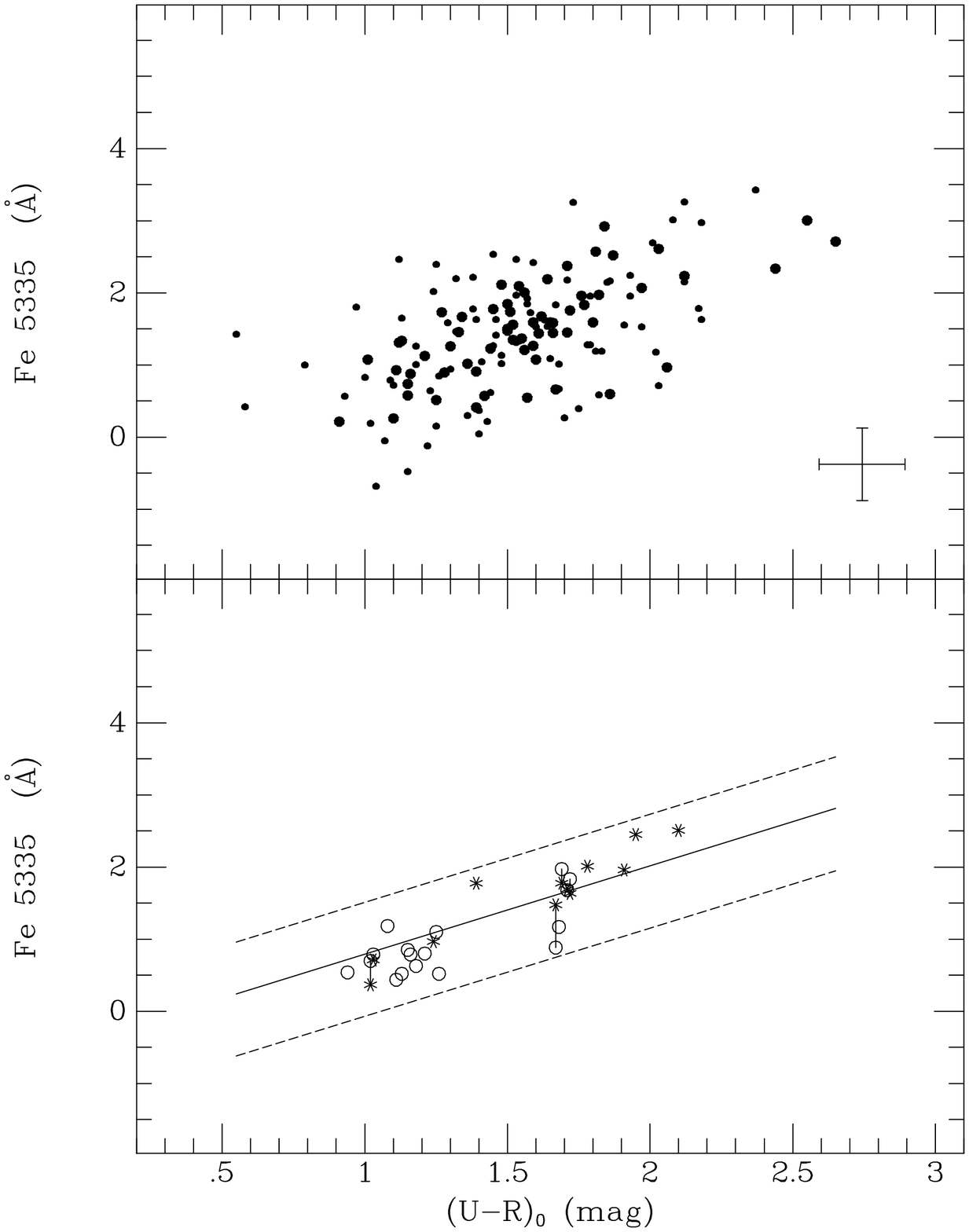]
{The same as Figure 2 for the Fe 5335 index.}
\label{fig5}

\figcaption[ 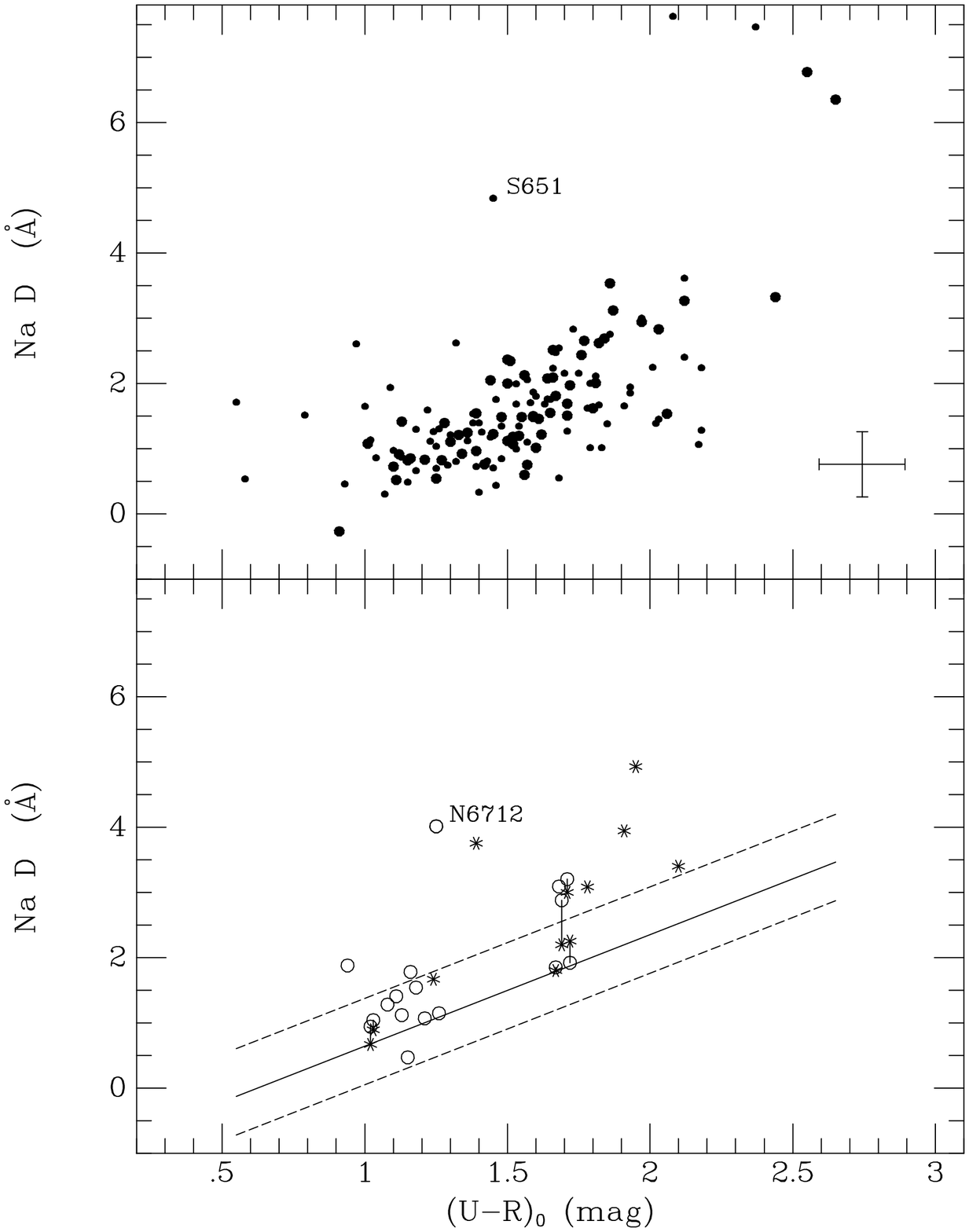]
{The same as Figure 2 for the NaD index.}
\label{fig6}

\figcaption[ 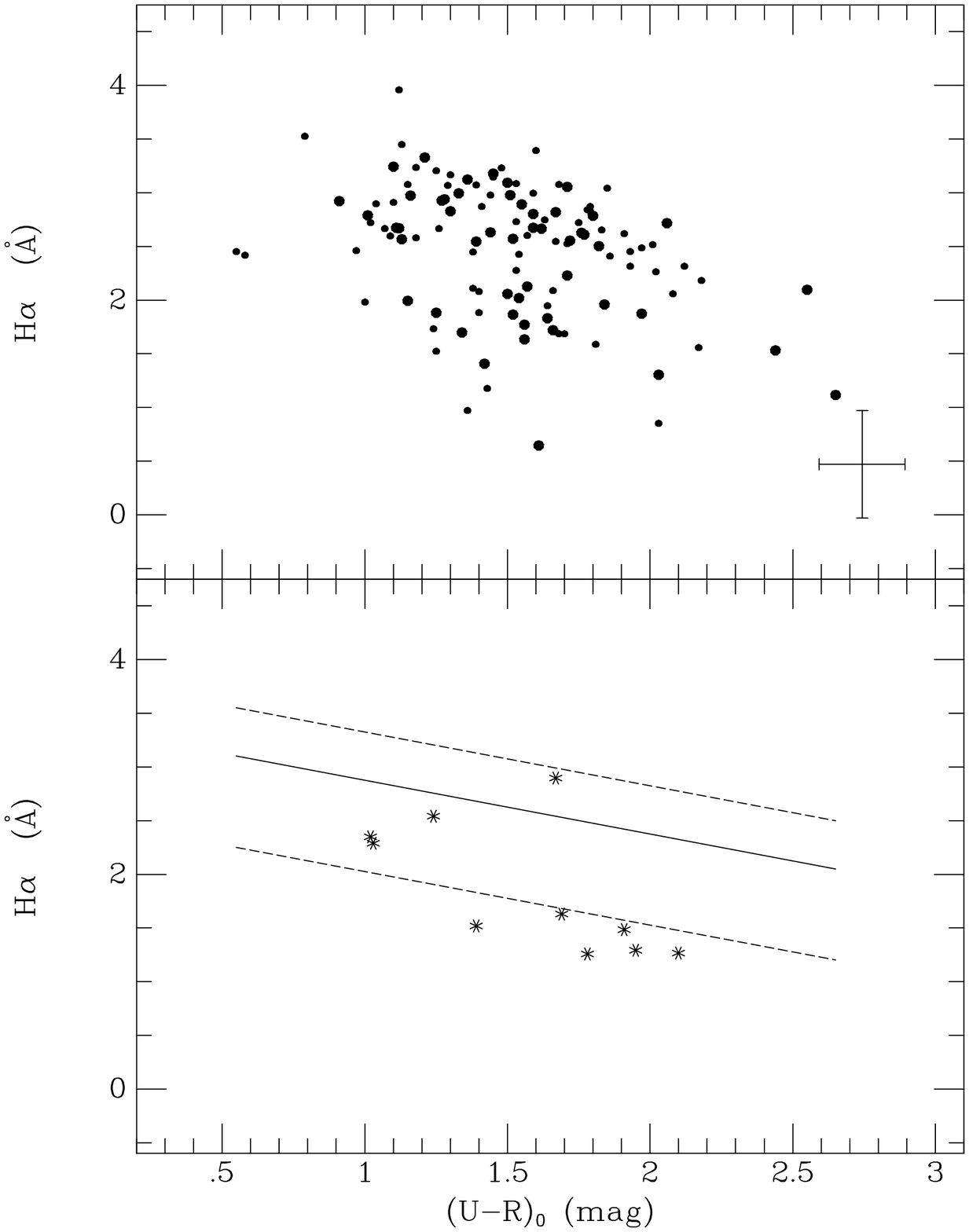]
{The same as Figure 2 for the $H_{\alpha}$ index.}
\label{fig7}

\figcaption[ 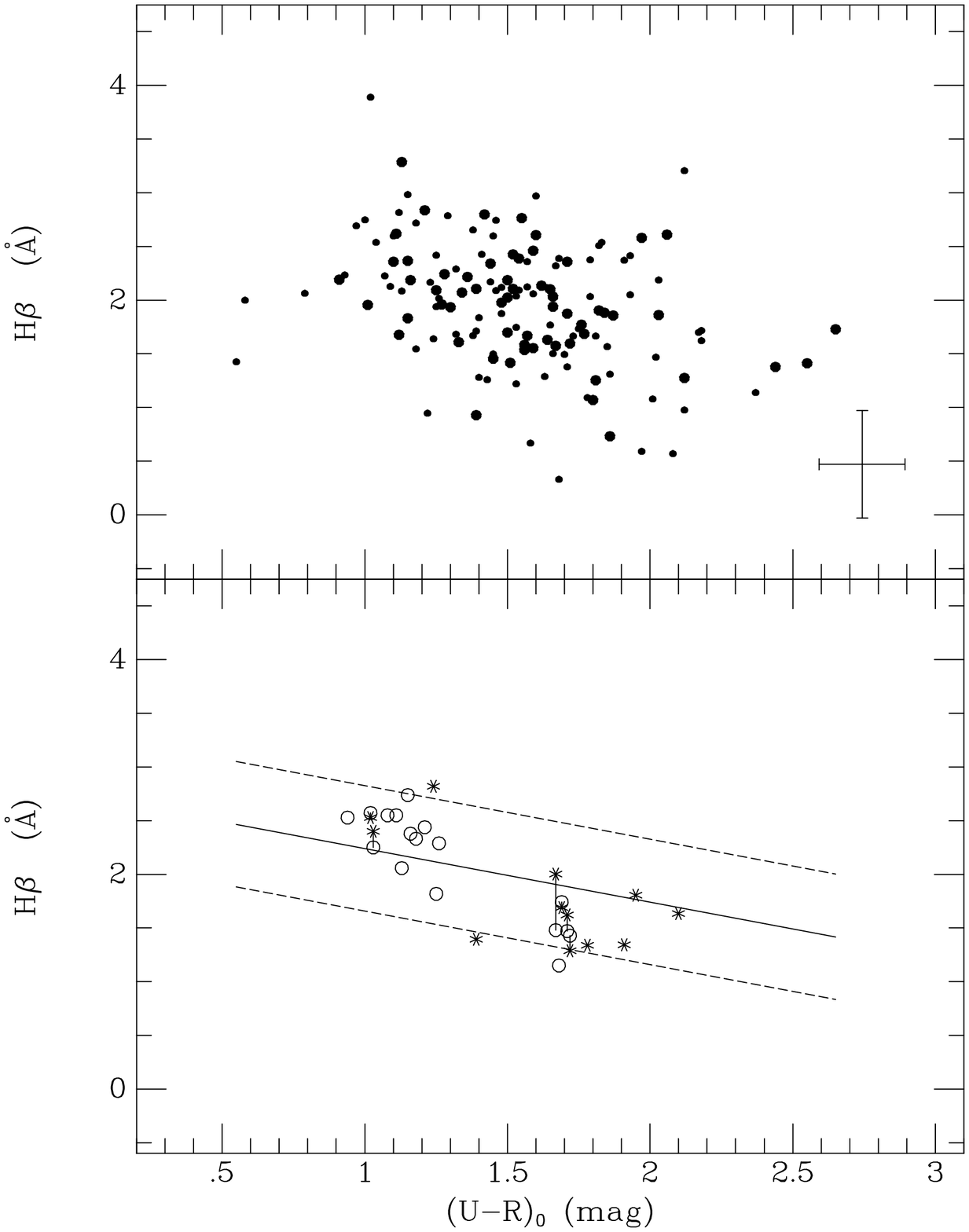]
{The same as Figure 2 for the $H_{\beta}$ index.}
\label{fig8}

\figcaption[ m87_fig9.ps]
{The NaD index is shown as a function of the Mg $b$ index.  The
upper panel shows the data for the M87 GCs, with
objects with $QSNR \ge 40$ indicated by the larger circles.  In the
lower panel the data for the galactic GCs (``*'' our data,
open circles from Burstein et al 1984) is compared with the
``median line'' of the M87 sample.}
\label{fig9}

\figcaption[ 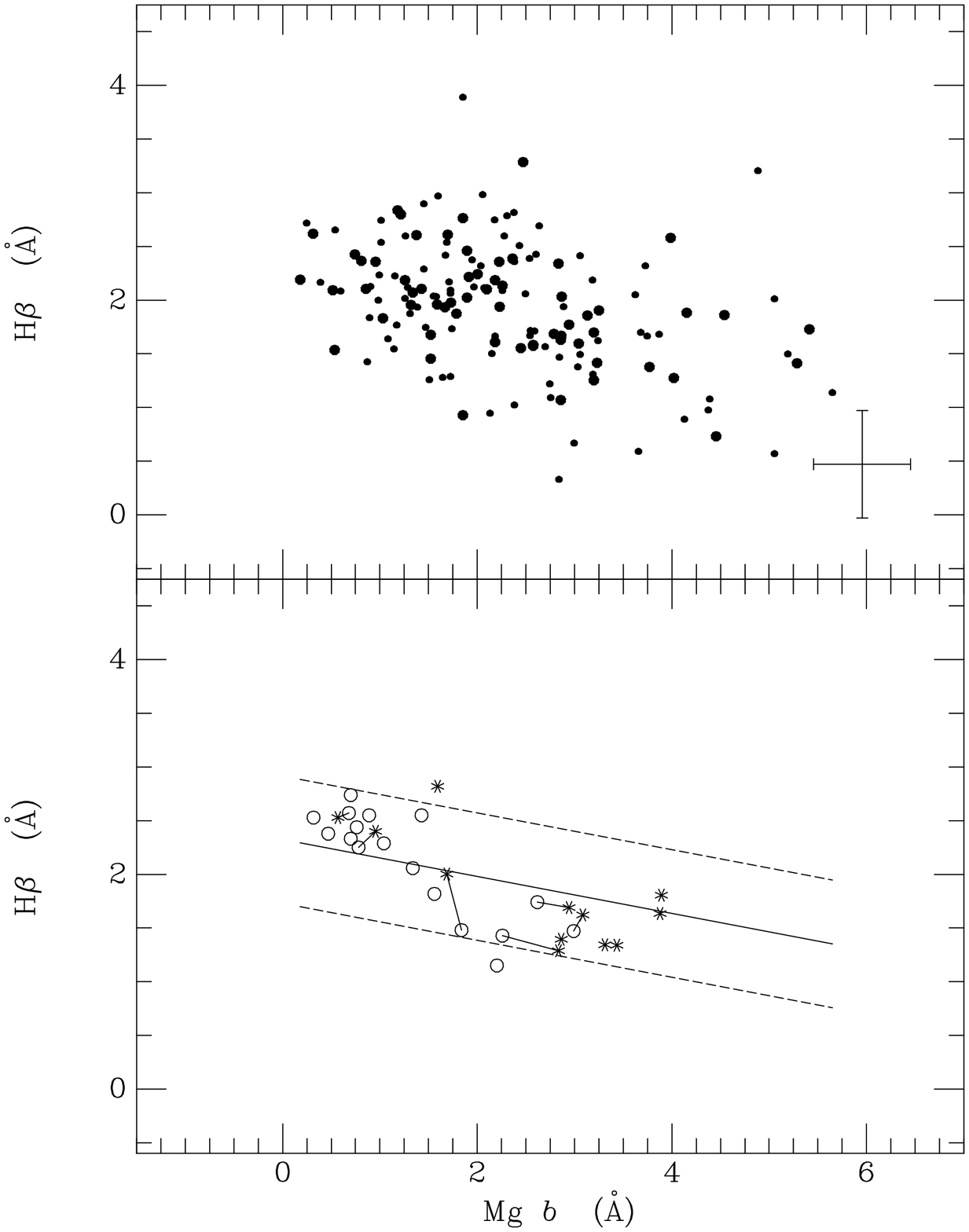]
{The same as Figure 9 for the $H_{\beta}$ index.}
\label{fig10}

\figcaption[ 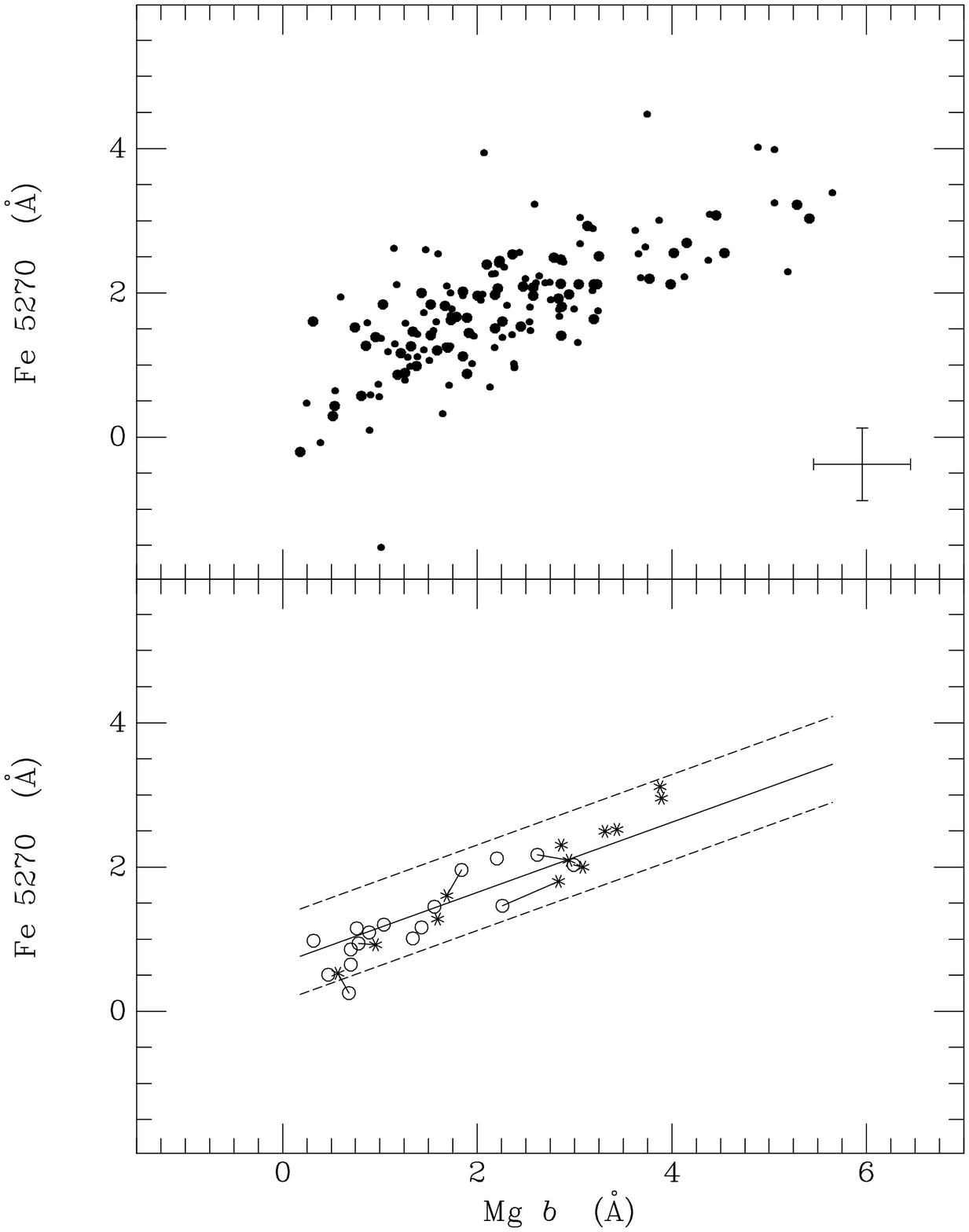]
{The same as Figure 9 for the Fe 5270 index.}
\label{fig11}

%

\figcaption[ 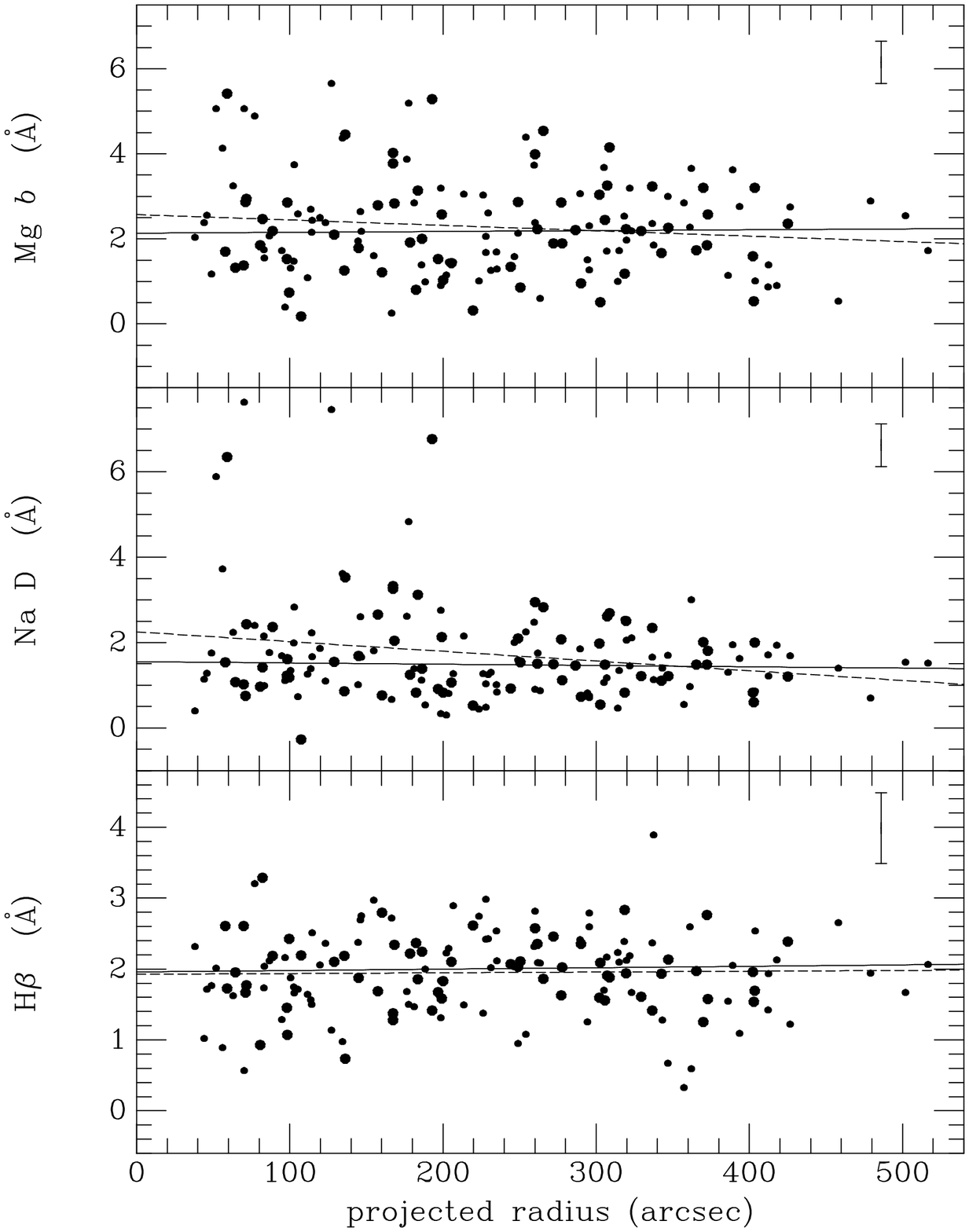]
{The Mg $b$, NaD and $H_{\beta}$ indices for the M87 GCs
are shown as a function of projected radius.  The least squares
fit for each index is indicated by a dashed line, while the
``median line'' is shown as a solid line.}
\label{fig12}

\figcaption[ 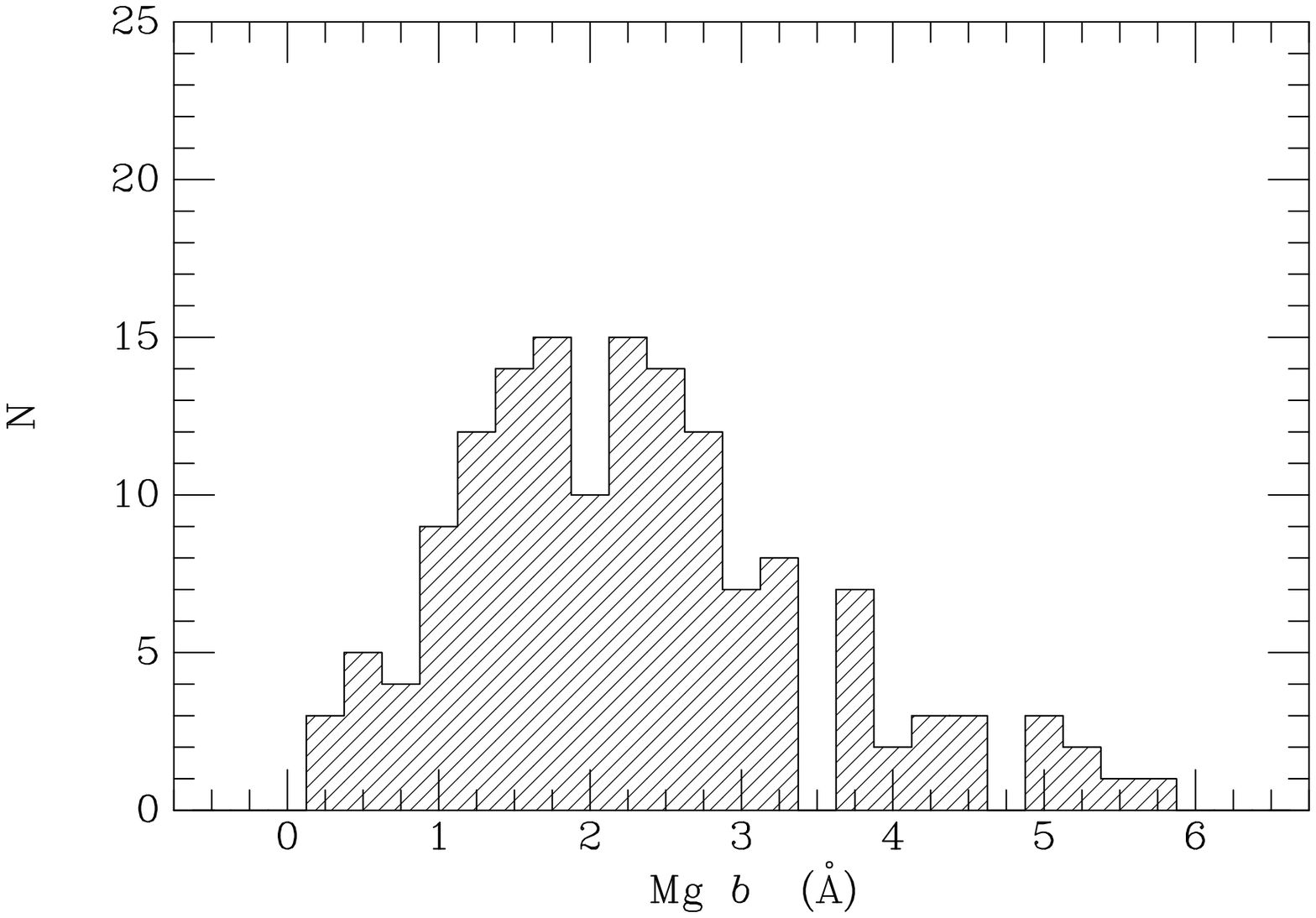]
{The histogram of the Mg $b$ indices for 150 M87 GCs
is shown.}
\label{fig13}

\figcaption[ 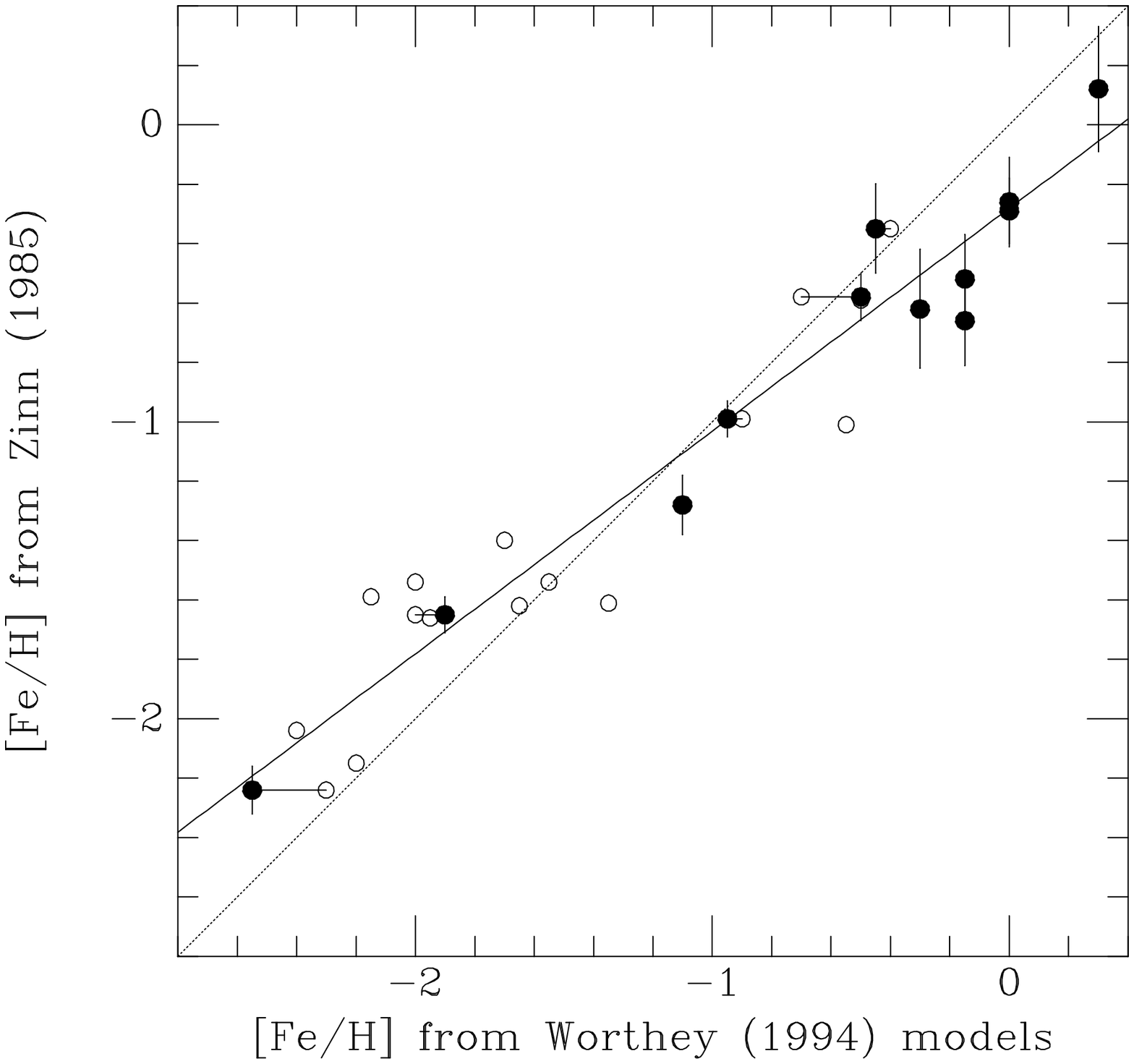]
{The metallicity from Zinn (1985) of the galactic globular 
clusters in our calibration
sample is shown as a function of that derived for them
using the Worthey (1994) models and their observed indices.}
\label{fig14}

\figcaption[ 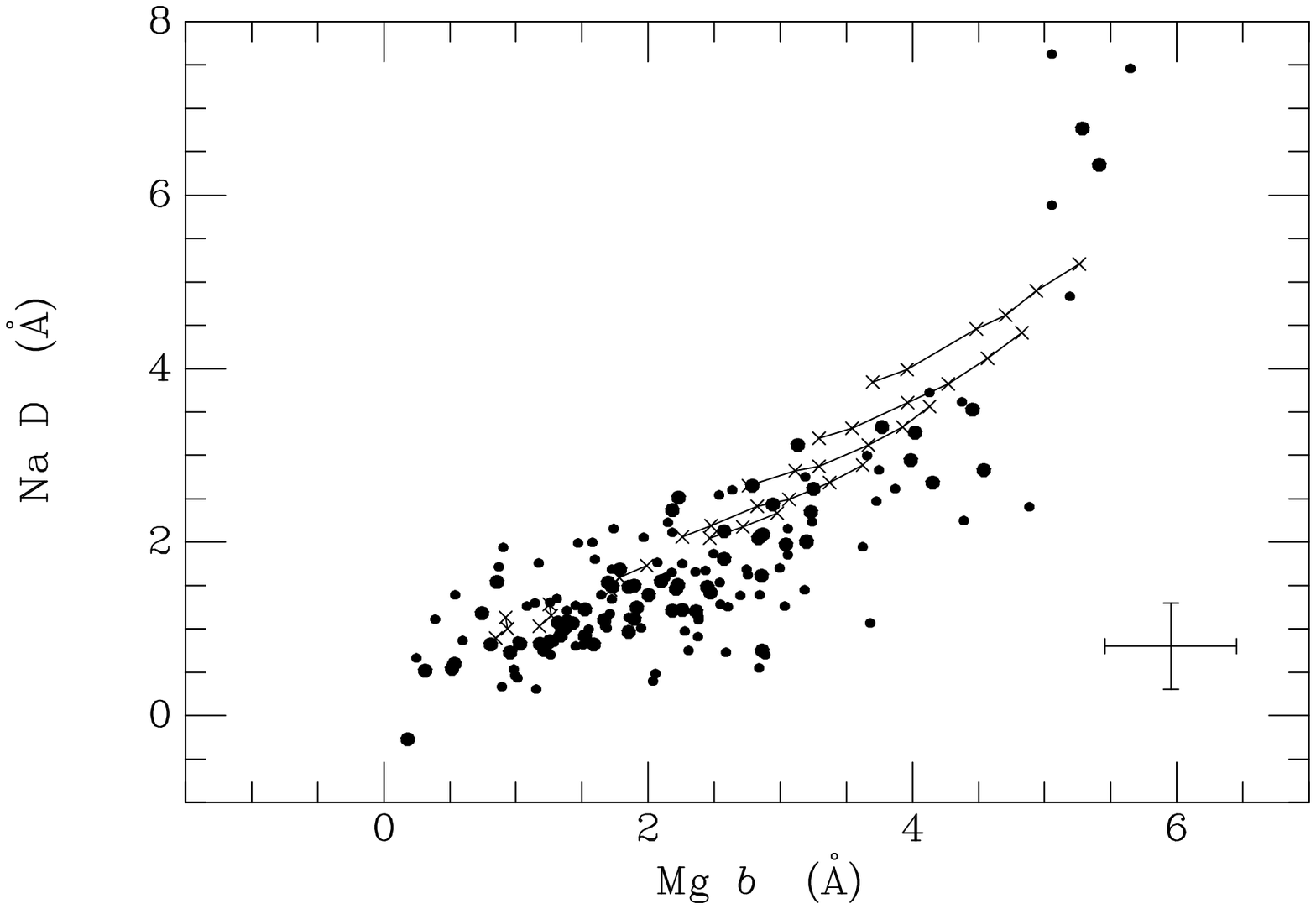]
{The NaD indices for 150 M87 GCs are shown as a function
of their Mg $b$ indices with the
predictions of the Worthey (1994) models overlaid.  Each line
connects models of fixed metallicity with varying ages.}
\label{fig15}

\figcaption[ 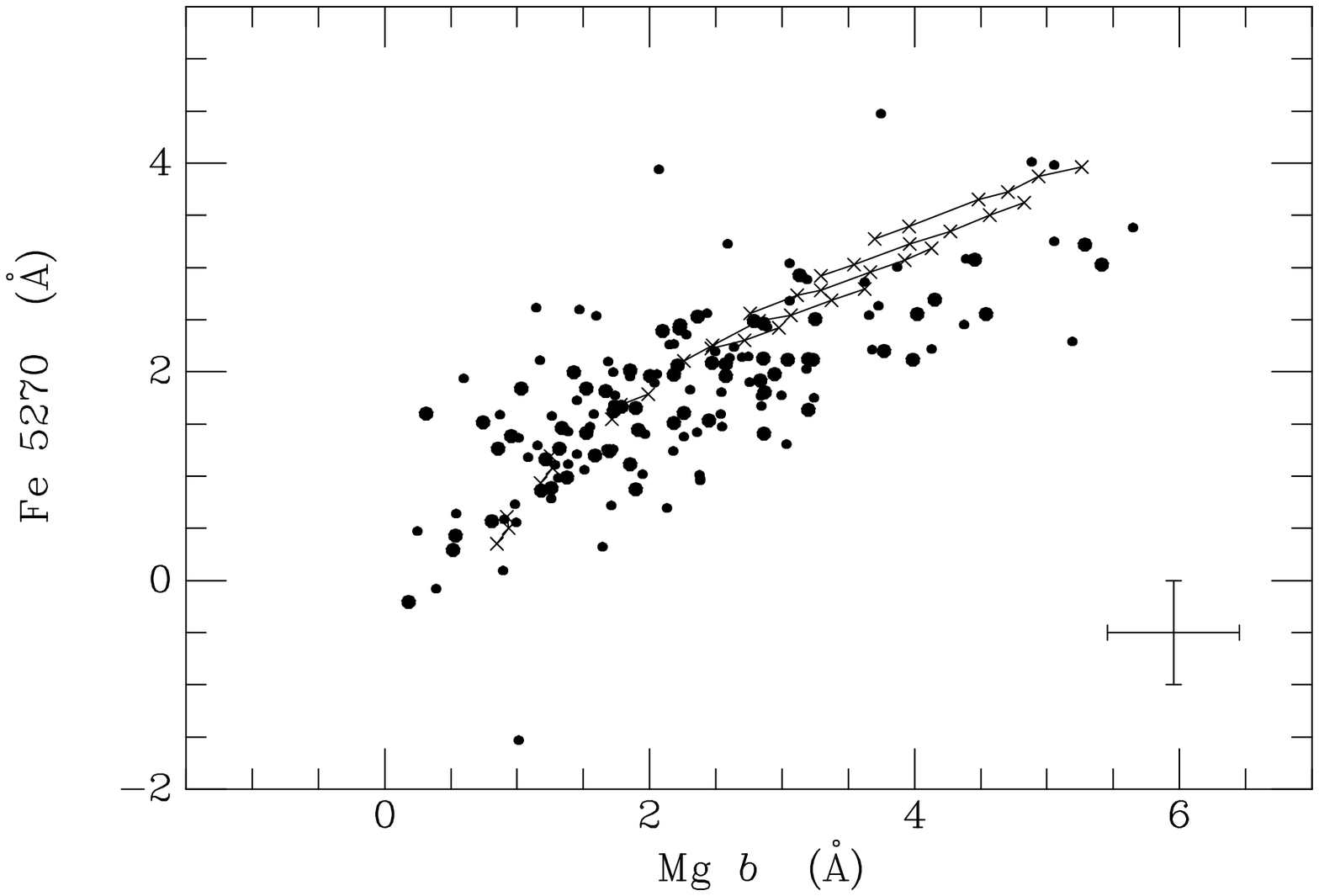]
{The same as Figure 15 for the Fe 5270 indices.}
\label{fig16}

\figcaption[ 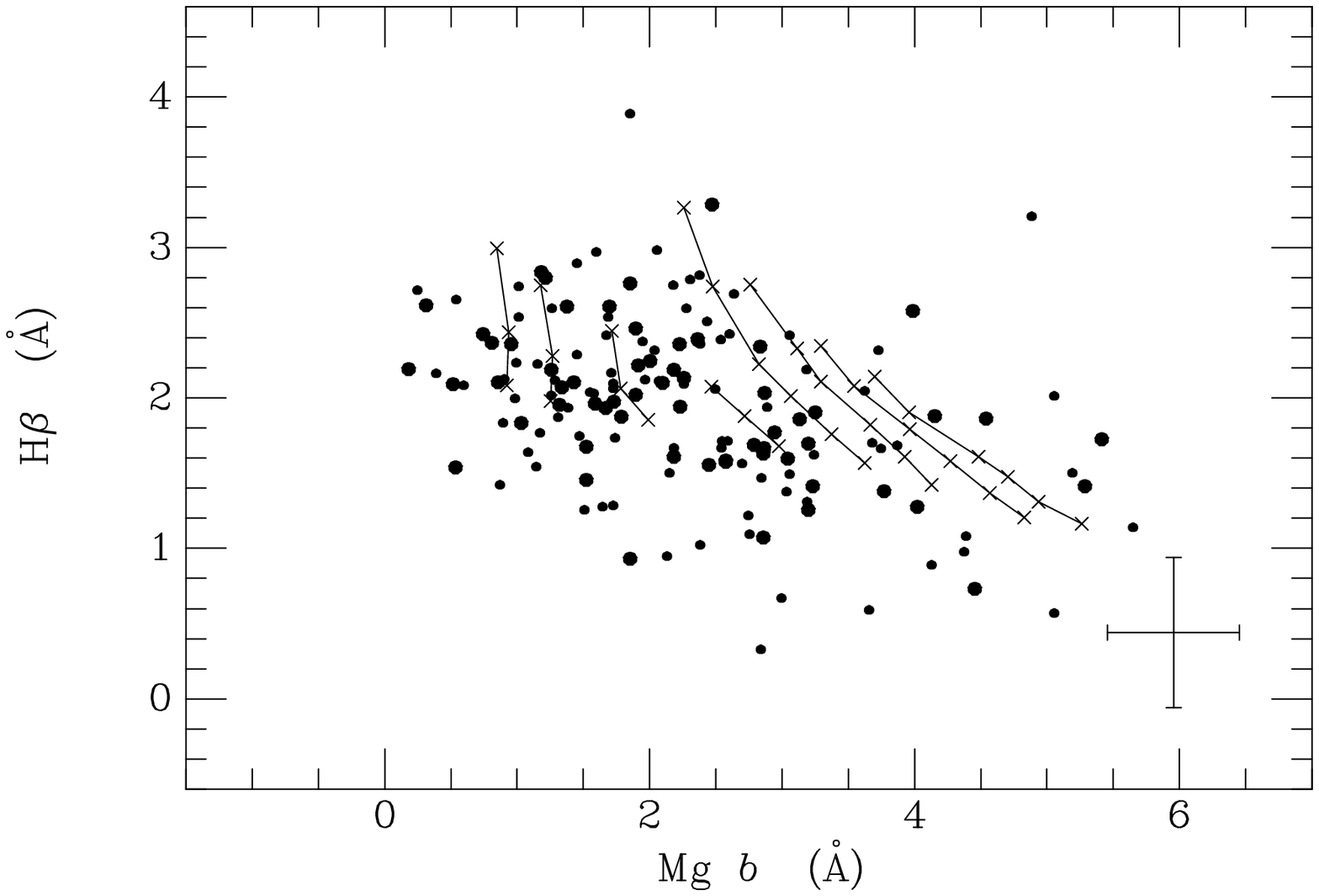]
{The same as Figure 15 for the $H_{\beta}$ indices.}
\label{fig17}

\figcaption[ 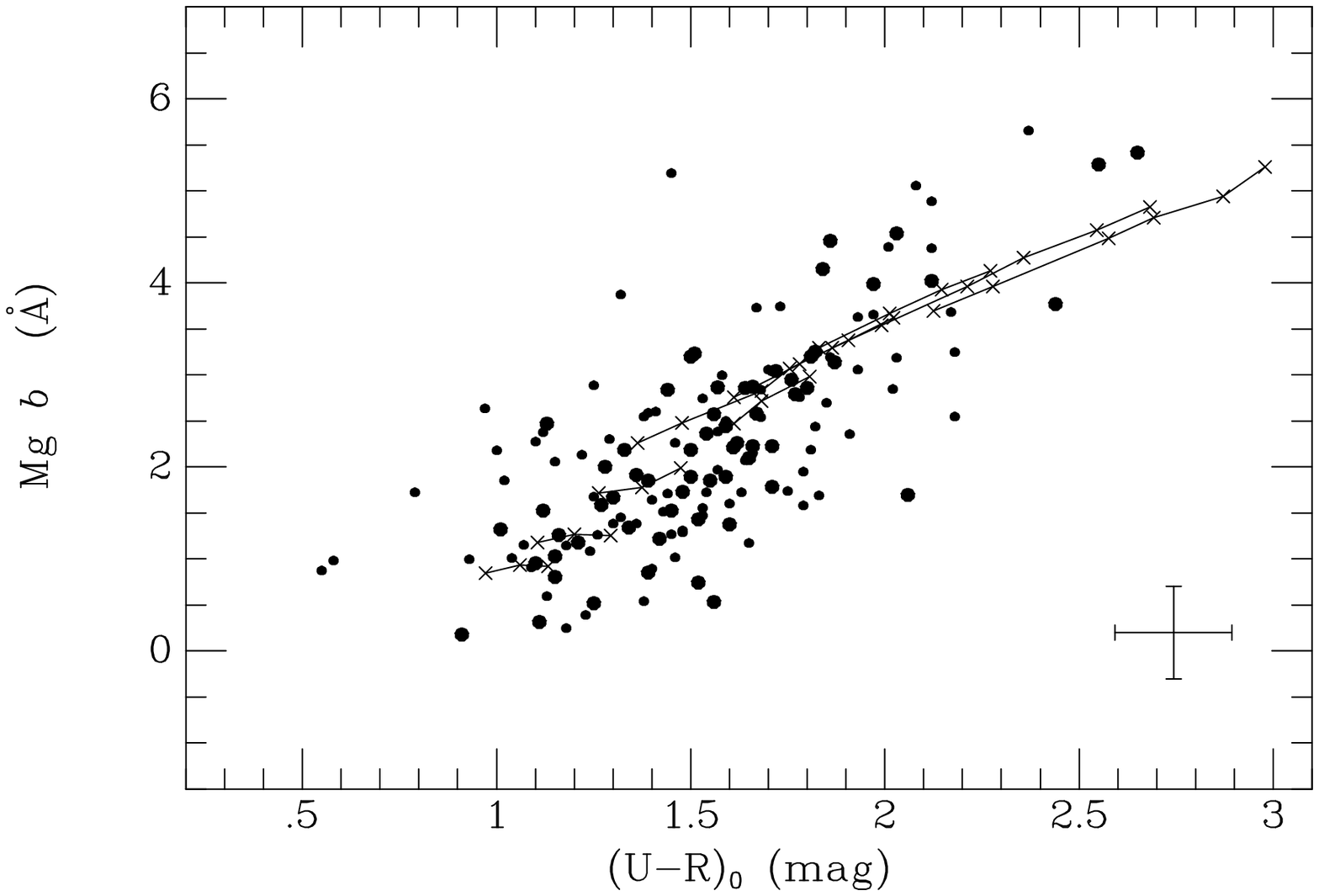]
{The Mg $b$ indices for the M87 GCs are shown as a
function of $(U-R)$ color.  The Worthey (1994) models
are overlaid as in Figure 15.}
\label{fig18}

\figcaption[ 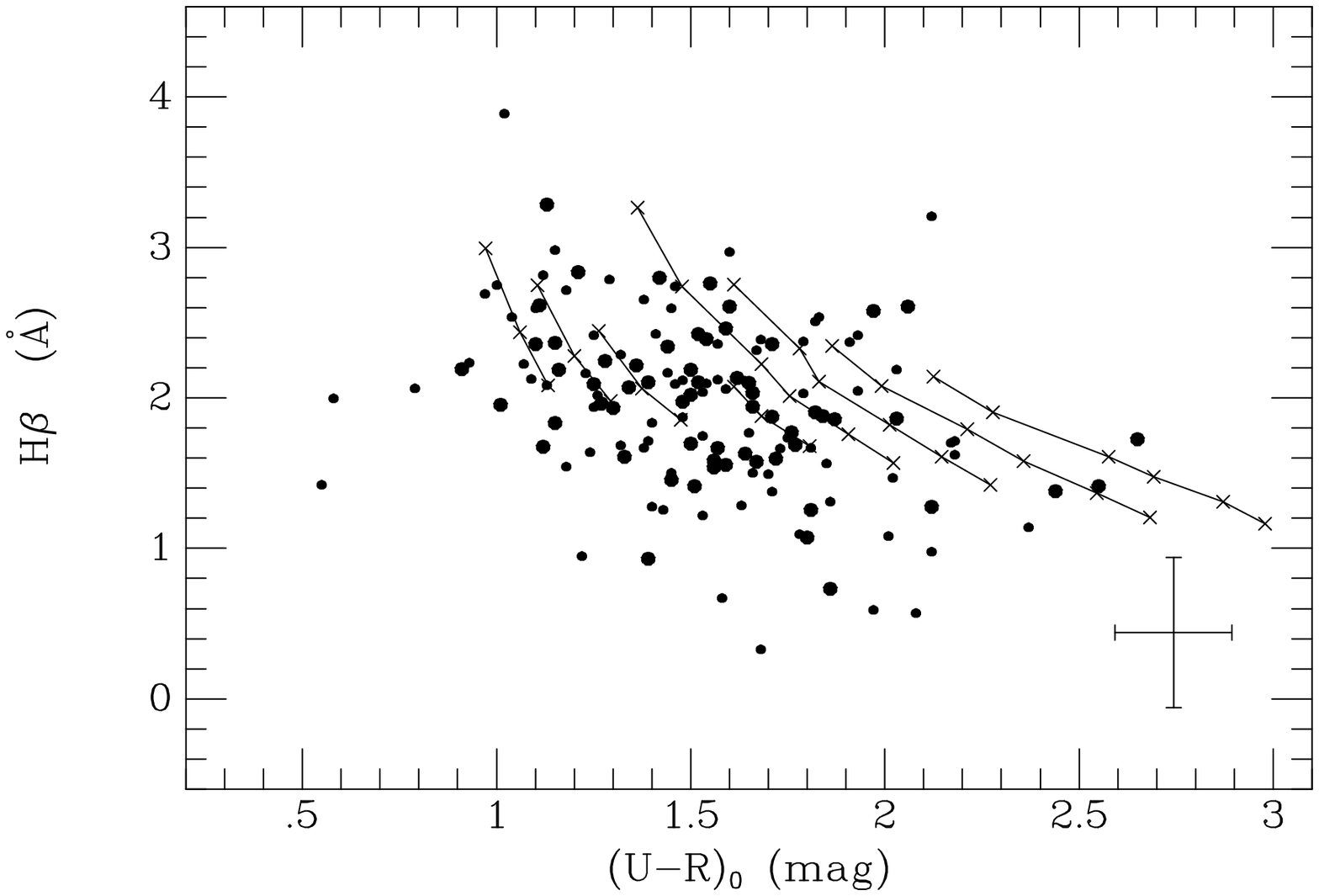]
{The same as Figure 18 for the $H_{\beta}$ indices.}
\label{fig19}

\figcaption[ 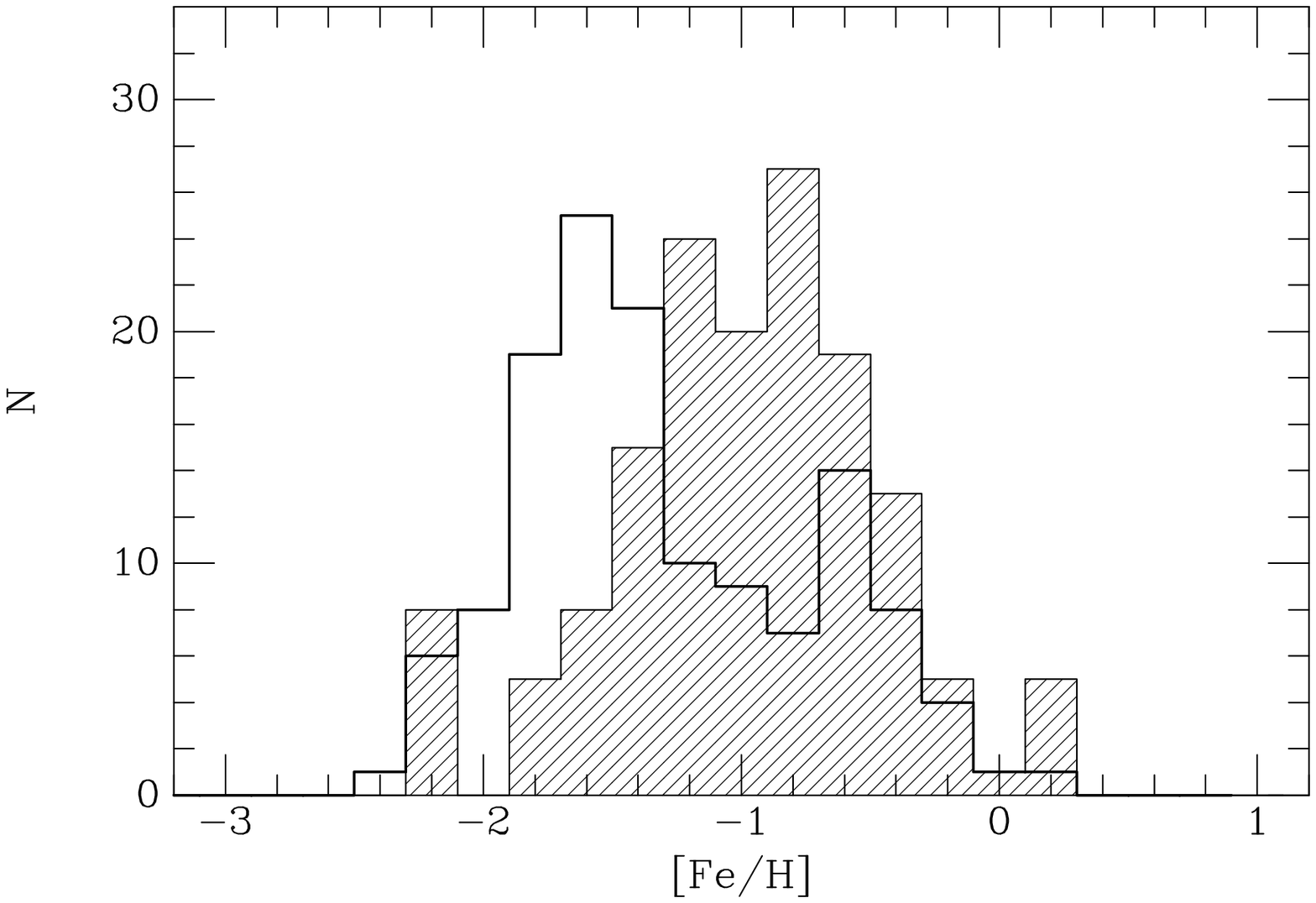]
{The histogram of metallicity for 150 M87 GCs is shown (hatched).
The solid line represents that of the galactic GC system
from the online compilation of W.~Harris.}
\label{fig20}

\figcaption[ 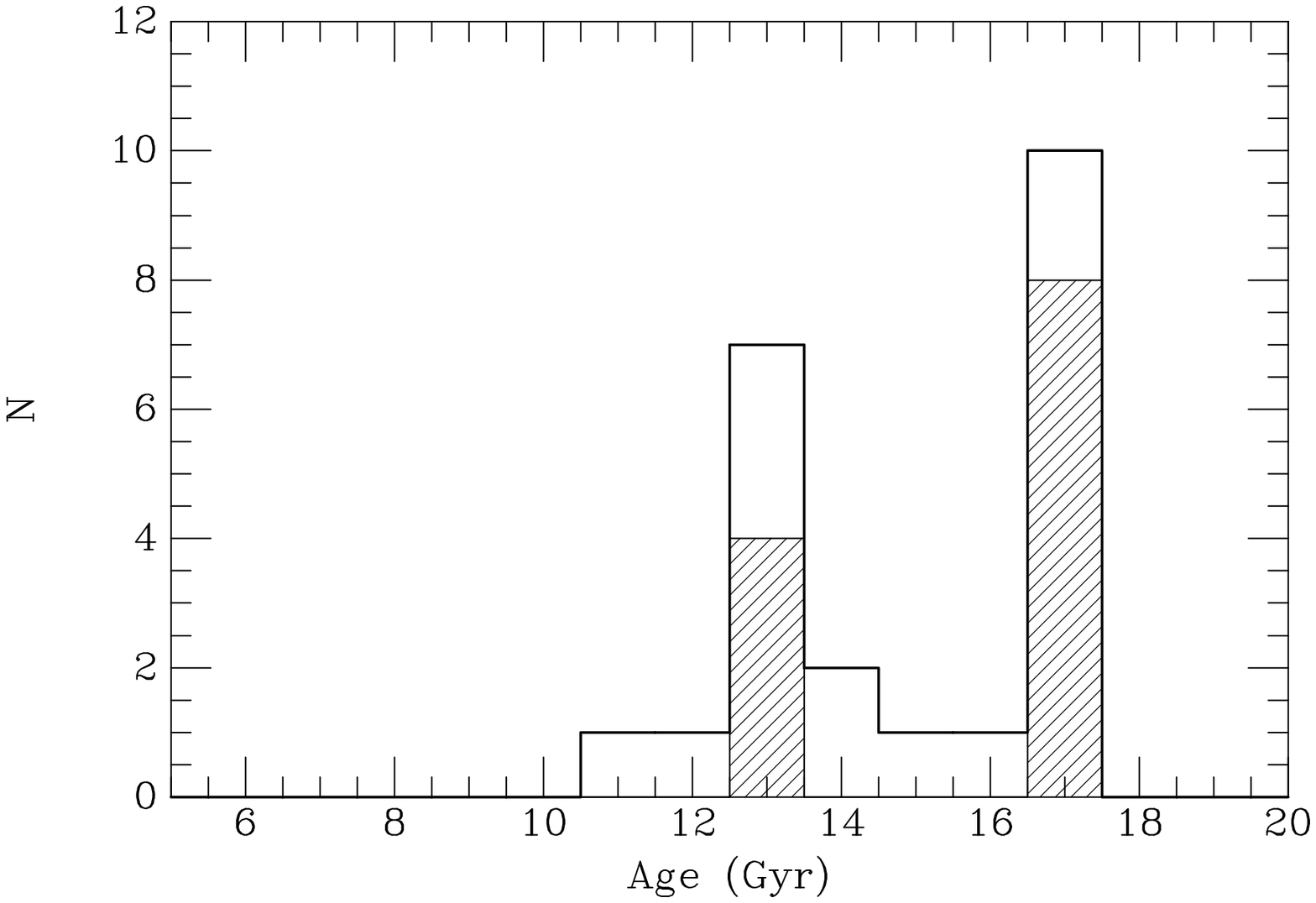]
{The histogram of ages derived for a sample of galactic globular
clusters using the Worthey (1994) models applied just to the data
presented here (hatched), and to our data combined with the 
sample of Burstein et al. (unhatched).}
\label{fig21}

\figcaption[ 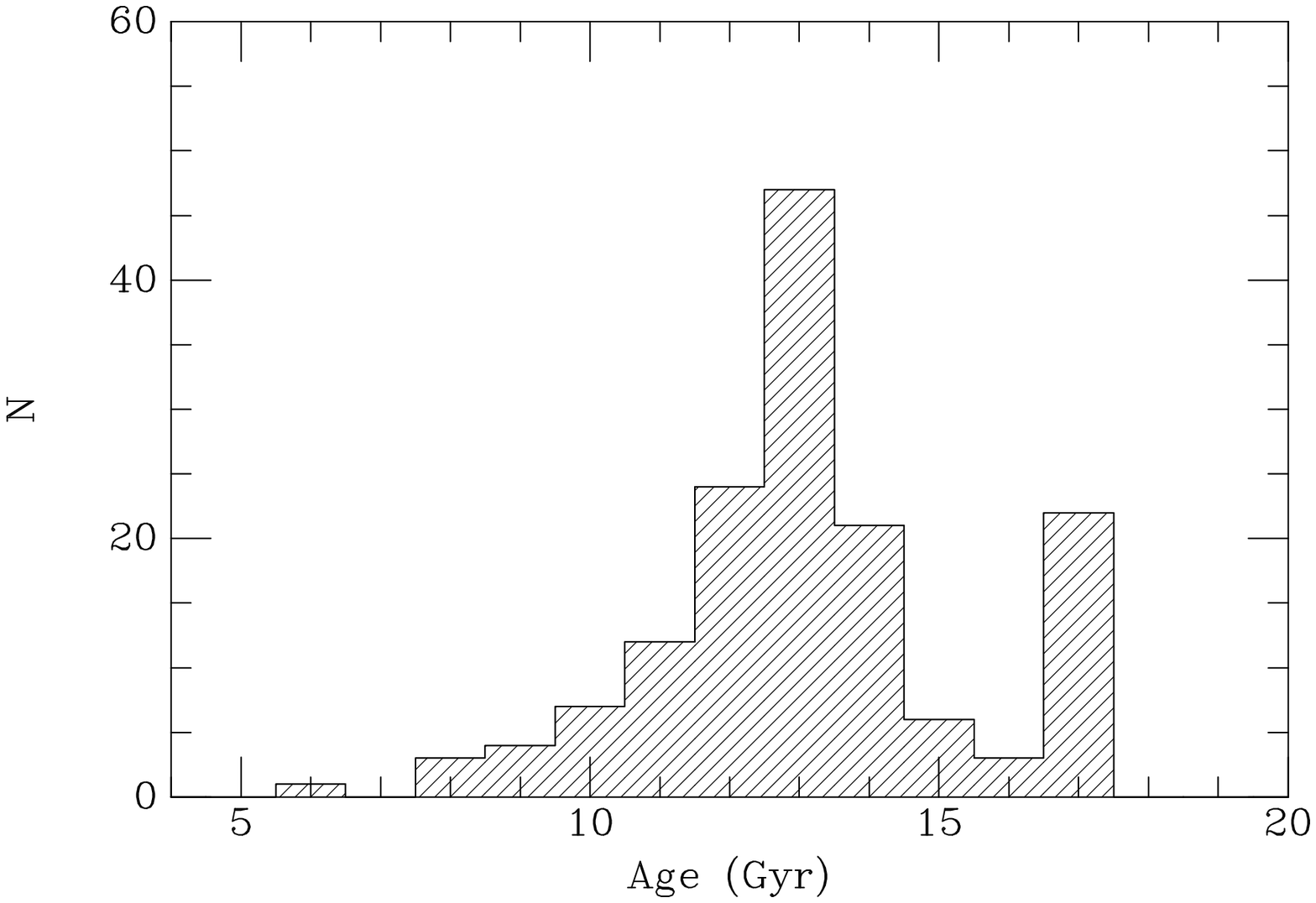]
{The histogram of ages derived for the M87 GC system
using the Worthey (1994) models applied to the data
presented here.}
\label{fig22}

\clearpage

\begin{figure}
\plotone{ m87_fig01.ps}
\end{figure}

\clearpage
\begin{figure}
\plotone{ m87_fig02.ps}
\end{figure}

\clearpage
\begin{figure}
\plotone{ m87_fig03.ps}
\end{figure}

\clearpage
\begin{figure}
\plotone{ m87_fig04.ps}
\end{figure}

\clearpage
\begin{figure}
\plotone{ m87_fig05.ps}
\end{figure}

\clearpage
\begin{figure}
\plotone{ m87_fig06.ps}
\end{figure}

\clearpage
\begin{figure}
\plotone{ m87_fig07.ps}
\end{figure}

\clearpage
\begin{figure}
\plotone{ m87_fig08.ps}
\end{figure}

\clearpage
\begin{figure}
\plotone{ 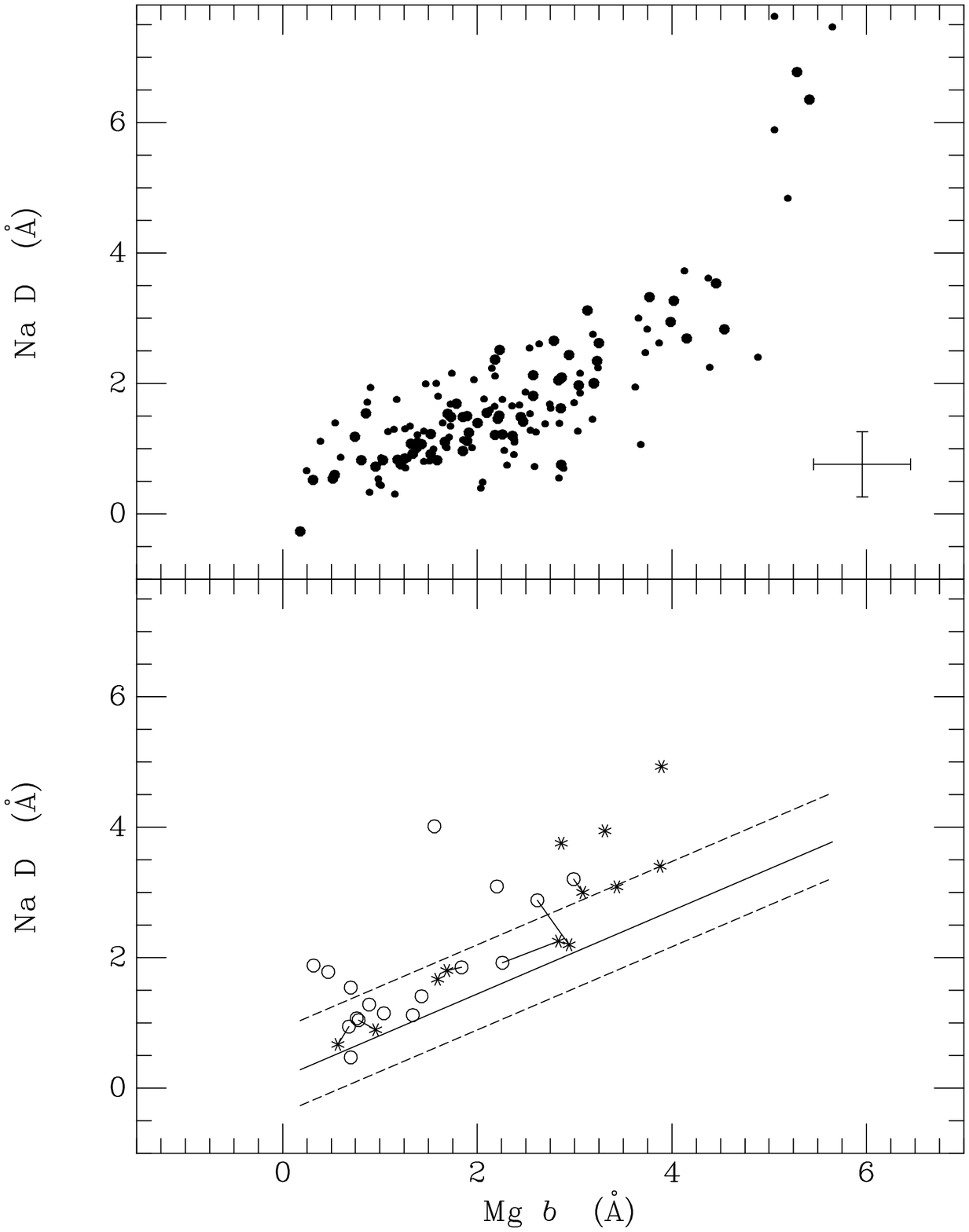}
\end{figure}

\clearpage
\begin{figure}
\plotone{ m87_fig10.ps}
\end{figure}

\clearpage
\begin{figure}
\plotone{ m87_fig11.ps}
\end{figure}

\clearpage
\begin{figure}
\plotone{ m87_fig12.ps}
\end{figure}

\clearpage
\begin{figure}
\plotone{ m87_fig13.ps}
\end{figure}

\clearpage
\begin{figure}
\plotone{ m87_fig14.ps}
\end{figure}
\clearpage

\begin{figure}
\plotone{ m87_fig15.ps}
\end{figure}

\clearpage
\begin{figure}
\plotone{ m87_fig16.ps}
\end{figure}

\clearpage
\begin{figure}
\plotone{ m87_fig17.ps}
\end{figure}
\clearpage
\begin{figure}
\plotone{ m87_fig18.ps}
\end{figure}

\clearpage
\begin{figure}
\plotone{ m87_fig19.ps}
\end{figure}

\clearpage
\begin{figure}
\plotone{ m87_fig20.ps}
\end{figure}

\clearpage
\begin{figure}
\plotone{ m87_fig21.ps}
\end{figure}

\clearpage
\begin{figure}
\plotone{ m87_fig22.ps}
\end{figure}


\clearpage
\begin{references}

\reference{} Angeletti, L. \& Giannone, P., 1997, A\&A, 321, 343
\reference{} Ashman, K.M., Bird, C.M., \& Zepf, S.E. 1994, AJ, 108, 2348
\reference{} Ashman, K.M. \& Zepf, S.E., 1992, ApJ, 384, 50
\reference{} Ashman, K.M. \& Zepf, S.E., 1997, Globular Cluster 
Systems (Cambridge: Cambridge University Press)
\reference{} Barnes, J.E. \& Hernquist, L., 1992, ARA\&A, 30, 705
\reference{} Bridges, T.J., Hanes, D.A. \& Harris, W.E., 1991, AJ, 101, 469
\reference{} Borges, A.C., Idiart, T.P., de Freitas Pacheco, J.A. 
\& Thevenin, F., 1995, AJ, 110, 2408
\reference{} Brodie, J.P. and Huchra, J.P., 1990, ApJ 362, 503
\reference{} Brodie, J.P. and Huchra, J.P., 1991, ApJ, 379, 157
\reference{} Burstein, D., Faber,S.M., Gaskell,C.M. and Krumm, N. 
1984, ApJ 287, 586
\reference{} Burstein, D. \& Heiles, C., 1984, ApJS, 54, 33
\reference{} Chavez, M., Malgnini, M.L. and Morossi, C., 1995, ApJ, 440, 210
\reference{} Cohen, J.G., 1975, ApJ, 197, 117
\reference{} Cohen, J.G., 1988, AJ, 95, 682
\reference{} Cohen, J.G., Djorgovski, S.G. \& McCarthy, J.K., 1997, 
manuscript in preparation
\reference{} Cohen, J.G. \& Ryzhov, A., 1997, ApJ (in press) (Paper~I)
\reference{} Couture, J., Harris, W.E. and Allbright,J.W.B., 1990, ApJS, 73, 671
\reference{} Dempster, A.P., Laird, N.M., \& Rubin, D.B. 1977, 
J.~R.\ Statist.\ Soc.\ B, 39, 1
\reference{} Dressler, A., 1993, in Neugebauer, G., ed., 
Palomar Observatory Annual Report 1993, 2
\reference{} Dressler, A., Lynden-Bell, D., Burstein, D., Davies, R.L.,
Faber, S.M., Terlevich, R.J. \& Wegner, G., 1987, ApJ, 313, 42
\reference{} Elson, R.A.W. \& Santiago, B.X., 1996a, MNRAS, 278, 617
\reference{} Elson, R.A.W. \& Santiago, B.X., 1996b, MNRAS, 280, 971
\reference{} Faber, S.M., Friel,E.D., Burstein, D. and Gaskell,C.M., 
1985, ApJS,57,711
\reference{} Fabian, A.C., Nulsen, P.E.J. \& Canizares, C.R., 1984, 
Nature, 310, 733
\reference{} Geisler, D. \& Forte, J.C., 1990, ApJ, 350, L5
\reference{} Gorgas, J., Faber, S.M., Burstein, D., Jesus Gonzalez, J. Courteau,S. and Prosser,C. 1993,ApJS, 86, 153
\reference{} Hanes, D.A. and Brodie, J.P., 1986, ApJ, 300, 279
\reference{} Harris, W.E., 1991,  ARA\&A, 29, 543
\reference{} Harris, W.E., 1996, AJ, 112, 1487
\reference{} Hobbs, L.M., 1974, ApJ, 191, 381
\reference{} Kissler-Patig, M., Brodie, J.P., Schroder, L.L., 
Forbes, D.A., Grillmair, C.J. \& Huchra, J.P., 1997, AJ, submitted
\reference{} Koyama, K., Takano, S., \& Tawara, Y., 1991, Nature, 350, 135
\reference{} Kurucz, R.L. 1993,  CD-ROM 13, {\it ATLAS}9 Stellar Atmosphere
Programs (Cambridge: Smithsonian Astrophysical Observatory) \par
\reference{} Lee, M.G. \& Geisler, D., 1993, AJ, 106, 493 
\reference{} Lu, L.M., Sargent, W.L.W., Barlow, T.A., Churchill, C.W. 
\& Vogt, S.S., ApJS, 107, 475
\reference{} Matsumoto, H., Koyama, K., Awaki, H., Tomida, H., Tsuru, T.,
Mushotzky, R. \& Hatsukade, I., 1996, PASJapan, 48, 201
\reference{}  McCarthy, J.K., Lennon, D.J., Venn, K.A., Kudritzski, R.P.,
Puls, J. \& Navarro, F., 1995, ApJ, 455, L155
\reference{}  McLachlan, G. J. \& Basford, K. E. 1988, Mixture Models: 
Inference and Applications to Clustering (Marcel Dekker, New York)
\reference{}  McWilliam, A., Preston, G.W., Sneden, C. \& Searle, L. 1995,
AJ, 109, 2757
\reference{} Monteverde, M.I., Herrero, A., Lennon, D.J. 
\& Kudritzki, R.P., 1997, ApJ, 474, L107 
\reference{} Mould, J.R., Oke, J.B. \& Nemec, J.M. 1987, AJ,  93, 53
\reference{} Mould, J.R., Oke, J.B., de Zeeuw, P.T., \& Nemec, J.M. 1990, AJ, 99, 1823
\reference{} Muzzio, J.C., 1987, PASP, 99, 614
\reference{} Oke, J.B.,  J.G.Cohen, M.Carr, J.Cromer, A.Dingizian, F.H.Harris,
S.Labrecque, R.Lucinio, W. Schaal, H.Epps \& J.Miller, 1995, PASP,
107, 307
\reference{} Ostrov, P., Geisler, D. \& Forte, J.C., 1993, AJ, 105, 1762
\reference{} Press, W.H., Flannery, B.P., Teukolsky, S.A. \& Vetterline, W.T., 1986,
	Numerical Recipes, Cambridge University Press
\reference{} Quinn, P. J. 1984, ApJ, 279, 596
\reference{} Rabin, D., 1982, ApJ, 261, 85
\reference{} Racine, R., Oke, J.B. and Searle, L. 1978, ApJ, 223, 82
\reference{} Reed, B.C., Hesser, J.E. and Shawl, S.J., 1988, 
PASP, 100, 545
\reference{} Rich, R.M., Sosin, C., Djorgovski, S.G., Piotto, G., King, I.R.,
Renzini, A., Phinney, E.S., Dorman, B., Liebert, J. \& Meylan, G., 1997,
ApJ, 484, L25
\reference{} Robinson, L. \& Wampler, E.J., 1972, PASP, 84, 161
\reference{} Strom, S.E., Forte, J., Harris, W., Strom, K.M., Wells, D., \& Smith, M. 1981, ApJ, 245, 416
\reference{} van den Bergh, S., 1975,  ARA\&A, 13, 217
\reference{} VandenBerg, D.A., Bolte, M. \& Stetson, P.B., 1996, 
Ann.Revs.Astr.Astrophys., 36,461
\reference{} Wheeler, J.C., Sneden, C. \& Truran, J.W. 1989,
Ann.Revs.Astr.Astrophys., 27, 279
\reference{} Whitmore, B.C., Sparks, W.B., Lucas, R.A., Duccio Mcchetto, F. 
\& Biretta, J.A., 1995, ApJ, 454, L73
\reference{} Worthey, G. 1994, ApJS, 95, 107
\reference{} Worthey, G., Faber, S.M. \& Gonzalez, J.J., 1992, ApJ, 398, 69
\reference{} Worthey, G. \& Ottaviani, D.L., 1997, ApJS, 111, 445
\reference{} Zinn, R., 1985, ApJ, 293, 424
\end{references}
\end{document}